 \documentclass[final,3p,times,fleqn]{elsarticle}

\usepackage{graphicx}
\usepackage{subcaption}
\usepackage{amssymb}
\usepackage{amsmath}
\usepackage{amsthm}
 
\DeclareMathOperator{\tr}{tr}

\usepackage{lineno}
\usepackage{bm}

\usepackage{color}
\usepackage{hyperref}

\journal{International Journal of Multiphase Flow}

\begin{document}

\begin{frontmatter}


\title{A two-fluid model for numerical simulation of shear-dominated suspension flows}


\author[Purdue]{Federico Municchi\fnref{fn1}}
\ead{Federico.Municchi@nottingham.ac.uk}
\ead[url]{https://github.com/fmuni}

\author[NITK]{Pranay P. Nagrani}

\author[Purdue]{Ivan C. Christov\corref{corrauth}}
\ead{christov@purdue.edu}
\ead[url]{http://tmnt-lab.org}

\address[Purdue]{School of Mechanical Engineering, Purdue University, West Lafayette, Indiana 47907, USA}
\address[NITK]{National Institute of Technology Karnataka, Surathkal, Mangalore  -- 575 025, Karnataka, India}

\fntext[fn1]{Present address: School of Mathematical Sciences, The University of Nottingham, University Park, Nottingham, NG7 2RD, UK.}
\cortext[corrauth]{Author to whom correspondence should be addressed.}

\begin{abstract}
Suspension flows are ubiquitous in nature (hemodynamics, subsurface fluid mechanics, etc.) and industrial applications (hydraulic fracturing, CO$_2$ storage, etc.). However, such flows are notoriously difficult to model due to the variety of fluid-particle and particle-particle interactions that can occur. In this work, we focus on non-Brownian shear-dominated suspensions, where kinetic collisions are negligible and frictional effects play a dominant role. Under these circumstances, irreversible phenomena such as particle diffusion and migration arise, requiring anisotropic stress models to describe the suspension rheology. On a continuum level, reduced-order models such as the suspension balance model (SBM) or the diffusive flux model are commonly used to predict particle migration phenomena. We propose a new method based on a two-fluid model (TFM), where both the phases are considered as interpenetrating continua with their own conservation of mass and momentum equations. Without employing the nowadays customary simplifications in applying the SBM, we close the ``full'' TFM instead. Specifically, we show that when an anisotropic stress analogous to that used in the SBM is added to the equilibrium equations for the particle phase, the TFM is able to accurately predict particle migration. Thus, the TFM does not require the assumptions of a steady suspension velocity and a Stokesian (inertialess) fluid, and the TFM can be easily extended to include buoyancy and even kinetic collisional models. We present several benchmark simulations of our TFM implementation in OpenFOAM{\textsuperscript\textregistered}, including in curvilinear coordinates and three-dimensional flow. Good agreement between the TFM solutions and previous experimental and numerical results is found.
\end{abstract}


\end{frontmatter}


\section{Introduction}
\label{S:I}

Non-Brownian suspensions are found in a wide range of applications, ranging from agriculture \cite{DC2015} to hydraulic fracturing \cite{TTK2019} and many more. Despite their ubiquitous presence in nature and engineering, understanding and modeling the physics of dense suspensions is still a frontier topic of modern fluid mechanics \cite{SP05,DM14}. In fact, while many problems related to dense granular flows are close to being understood \cite{jfp06,fp08}, a full tensorial form for the suspension stress, which can be applied to any geometry, is not yet available \citep{GP18,M2016}.

\subsection{Shear flows of suspensions}
\label{SS:I.1}

Shear-dominated suspensions play a significant role in industrial processes, for example those involving particle separation \cite{Strathmann2001}. This is because of the ``peculiar'' irreversible phenomena, such as shear-induced particle migration and particle diffusion \cite{la87}, that arise in the flow of suspensions in this regime. Shear-induced particle migration  leads to a net particle flux from regions of high shear rate to regions of low shear rate, and it was  believed to originate from long-range hydrodynamic interactions between particles \cite{la87}, even if recent studies suggest particle collisions also play a role \cite{PMB15}. Such drift has obvious significant consequences in processes involving channel flows, since it tends to focus the particles near the channel's centerline. The phenomenon of self-diffusion in shear flows of suspensions was observed experimentally by Leighton and Acrivos \cite{la87} and was considered to be caused by the inhomogeneous particle distribution resulting from the migration flux. Their measurements suggested that these displacements alter the apparent suspension viscosity.

For the present study, the typical conditions encountered in shear-dominated suspension flows can be summarized as follows:
\begin{itemize}
    \item[(i)] the particles are large;
    \item[(ii)] the fluid flow is inertialess.
\end{itemize}

Condition (i) is generally quantitatively assessed by requiring that the particle diameter $d_p$ is larger that $\approx 1 ~\mu \text{m}$. Physically, this condition means that fluctuations of the flow field should not affect the particle phase significantly, and that effects of Brownian motion can be neglected. More precisely, the particle motion is mostly driven by the shear flow. Thus, condition (i) can also be quantified by introducing the particle P\'eclet number
\begin{equation}
    \label{eq::peclet_cond}
    Pe_p = \frac{ \dot{\gamma}d_p^2}{2D_p} \gg 1.
\end{equation}
Here, $\dot{\gamma}$ is a characteristic shear rate arising from the suspension flow, and $D_p$ is the characteristic Brownian diffusivity of the particles.

Condition (ii) requires that the particle Reynolds number $Re_p$ is small:
\begin{equation}
\label{eq::Rep_cond}
    Re_p = \frac{\rho_f U_f d_p}{\mu_f} \ll 1,
\end{equation}
where $\rho_f$ and $\mu_f$ are, respectively, the density and viscosity of the suspending fluid, and $U_f$ is a characteristic fluid velocity. In fact, Han et al.~\cite{HKK1999} observed that when the particle Reynolds number is increased above a certain threshold ($Re_p \approx 0.2$ for tubes) particle migration is replaced by a different kind of irreversible mechanism: the Segr\'{e}--Silberberg effect \cite{SS1961}. Furthermore, Picano et al.~\cite{Picano2013} demonstrated that inertial shear thickening appears in shear-dominated dense particle suspensions. This effect is not accounted for in the rheological models that we consider, and thus we will always require that $Re_p$ is well below the threshold at which these effects must be taken into account.     

Energy applications such hydraulic fracturing (see, e.g., \cite{Hyman2016}) make use of shear-dominated dense suspensions. Specifically, suspensions of particles (termed `proppants') are injected into newly created fractures to increase fracture conductivity and prevent closure of the fracture upon the cessation of flow \cite{VE95}. The proppant distribution is thus critical in such applications. It has been argued that proppant transport involves precisely a dense suspension flows in the above-described regime of $Re_p\ll1$ and $Pe_p\gg1$ \cite{DP14}. Two Fluid Models (TFMs) are becoming popular for such applications \cite{DP14,DP15,SM16}, however, these are often further reduced to simplified sets of equations to allow for their numerical solution. Thus, a general numerical TFM framework that can be used to address proppant transport questions is lacking.

Finally, suspension flows do not have to be restricted to the regimes (and conditions) mentioned here. In general, a wide variety of transitional phenomena may occur. Therefore, methods for the numerical simulation of suspension flows should be flexible enough to account for flow with inertial and non-inertial regions, or flows in which kinetic collisions play an important role. We posit that TFMs could pave the way towards such general formulations.

\subsection{Numerical simulation of shear-dominated suspensions}

A wide range of numerical methods have been used to study the rheology of shear-dominated suspensions, and to predict the macroscopic behaviour of suspension flows. Maxey \cite{M17} reviews several discrete methods in which particles are tracked or accounted for individually. Such methods are generally employed to investigate the suspension rheology, as done by Yeo and Maxey \cite{YM2011}, who used the Force Coupling Method (FCM) to calculate normal stress differences of the suspension. However, these methods are too computationally expensive to be employed in the prediction of flows occurring in nature or during industrial processes. Thus, it is necessary to develop continuum models for suspensions.

Typically, these multiphase continuum models can be classified into three categories: Diffusive Flux Models (DFMs), Suspension Balance Models (SBMs) and Two Fluid Models (TFMs). DFMs are mixture models, in which the suspension is described as a single non-Newtonian fluid (i.e., the suspension is seen a single continuum \cite{GP18}) whose rheology depends upon the particle volume fraction. Although SBMs were originally derived based on an argument stemming from phase-averaging \cite{NB94,L09,Nott2011}, their most common application today is based on assuming a steady suspension velocity and reducing the problem to a single set of conservation equations with appropriate closures \cite{GuaMor,DLLM2013,M2016}. This approach allows one to derive expressions for the phenomenological closures in the DFMs. Thus, DFMs and ``standard'' SBMs require the solution of a vector parabolic equation for the mixture velocity and a scalar hyperbolic equation for the particle volume fraction, as well as the Poisson equation for the pressure. In these DFM and SBM models, an additional diffusive flux $\bm{J}_\mathrm{diff}$ arises in the advection equation for the particle volume fraction.

To make this distinction more clear, consider the  suspension's volume-averaged velocity
\begin{equation}
    \label{eq::suspension_U}
    \bm{u}_\mathrm{mix} = \phi \bm{u}_p + (1-\phi) \bm{u}_f,
\end{equation}
which is defined as the convex combination of the fluid velocity $\bm{u}_f$ and the particle velocity $\bm{u}_p$, with the particle volume fraction $\phi$ used as the weight. Then, the advection equation for $\phi$ becomes:
\begin{equation}
    \label{eq::adv_phi}
    \frac{\partial \phi }{\partial t} = - \nabla \cdot \left(  \bm{u}_p \phi \right) = - \nabla \cdot \left(  \bm{u}_\mathrm{mix} \phi \right) - \nabla \cdot \left[ \left(  \bm{u}_p - \bm{u}_\mathrm{mix} \right) \phi \right] = - \nabla \cdot \left( \bm{u}_\mathrm{mix} \phi \right) - \nabla \cdot \bm{J}_\mathrm{diff},
\end{equation}
having introduced the diffusive flux $\bm{J}_\mathrm{diff}$ as a closure accounting for the \textit{a priori} unknown flux arising due to the term $\left( \bm{u}_p - \bm{u}_\mathrm{mix} \right) \phi$. In the DFM \cite{la87,pabga92,vsb10}, this closure is fundamentally a phenomenological one, and the diffusive flux is modeled as:
\begin{equation}
    \label{eq::diffFlux}
    \bm{J}_{\mathrm{diff}} = - \left( D_{\phi} \nabla \phi + D_{\dot{\gamma}} \nabla \dot{\gamma}  \right).
\end{equation}
Here, $D_{\phi}$ and $D_{\dot{\gamma}}$ are diffusion coefficients determined empirically or through experiments. On the other hand, in the ``standard'' implementation of the SBM \cite{NB94,MB1999,MM2006,L07,Nott2011,DLLM2013} the diffusive flux is computed by upscaling the \emph{steady-state} equations of a two fluid model. As result, $\bm{J}_\mathrm{diff}$ is given by the product of the particle mobility $M$ with the divergence of the particle phase stress tensor $\bm{\Sigma}_p$:
\begin{equation}
    \label{eq::SBM}
    \bm{J}_\mathrm{diff} =  - M \nabla \cdot \bm{\Sigma}_p.
\end{equation}
Vollebregt et al.~\cite{vsb10} showed that the DFM and these SBM models are equivalent in the case of isotropic particle phase stress, in which case $\nabla \cdot \bm{\Sigma}_p$ can be expressed as the gradient of a potential $\mu^{\star}$.

\subsection{The two fluid model}

The latter idea of expressing $\nabla \cdot \bm{\Sigma}_p$ as the gradient of a potential was recently employed by Drijer et al.~\cite{DLVS2018} to develop a TFM using the computational environment provided in STAR-CCM+. In a TFM, the fluid and particles are modeled as two distinct interpenetrating continua, thus allowing for a more detailed description of the multiphase flow dynamics with respect to mixture models. However, this approach comes at a significant computational cost (compared to DFM and SBM), since two coupled vector equations have to be solved for the phases' velocity fields. In the formulation of Drijer et al.~\cite{DLVS2018}, an additional forcing term $\bm{F}_\mathrm{SID} = -\nabla \mu^{\star}$ is added to the particle momentum equation and subtracted from the fluid momentum equation. In their model, they defined $ \nabla \mu^{\star} =  \left( D_{\phi} \nabla \phi + D_{\dot{\gamma}} \nabla \dot{\gamma}  \right)/M$, which makes their model a combination of a DFM and a TFM.  However, this model cannot account for the anisotropy of the particle stress tensor, therefore it is not suitable for problems in curvilinear geometries.

It is crucial to note that while the SBM (in its current form widely used in the literature) and the DFM are indeed essentially multiphase flow models, they require the solution of only one momentum conservation equation for the whole suspension. Therefore, the suspension is modeled as a single fluid whose rheology depends on the scalar field $\phi$ (the particle volume concentration). Therefore, a total of five coupled equations (three for $\bm{u}_\mathrm{mix}$, one for the hydrodynamic pressure $p$, and one for $\phi$) have to be solved. By contrast, a TFM requires the solution of the momentum conservation equations for both phases, leading to a total of \emph{eight} coupled equations (three for $\bm{u}_p$, three for $\bm{u}_f$, one for $p$ and one for $\phi$). This is the fundamental mathematical, computational and physical difference between the proposed TFM and previous models employed for the simulation of dense non-Brownian suspensions.

The TFM formulation of multiphase flows is very popular in the particulate flow community, since it is able to accurately capture non-equilibrium phenomena, which is an essential feature for developing upscaled models like the filtered TFM \cite{MMHAS13}. On the other hand, the SBM and DFM formulation are mostly employed in the suspensions community. A notable exception is that, recently, Ahnert et al.~\cite{AMW19} used a TFM to solve for certain plane Poiseuille flows of suspensions.

\subsection{Goals and outline of the paper}
\label{SS:I.4}

In this work, we aim to establish a formulation of the TFM that is valid for shear-dominated suspension flows in general curvilinear geometries and that can be straightforwardly extended to collisional or inertial flows. To this end, we modify the \emph{twoPhaseEulerFoam} solver from the finite volume library OpenFOAM{\textsuperscript\textregistered} to include the anisotropic particle stress tensor models employed in the SBM. In our TFM formulation, particle migration is not modeled as an additional forcing term, rather it is incorporated in the particle phase stress. The objective is to provide an open-source implementation of a sufficiently general and extensible TFM that can be used for testing future rheological models or for specific applications of suspension flows in the limit of $Pe_p \gg 1$. 

To this end, this paper is structured as follows: in section \ref{S:M}, we describe the governing equations and the rheological models/closures that we employ. Section \ref{S:N} briefly outlines the numerical implementation of the method, with an emphasis on the anisotropic stress tensor. We demonstrate the accuracy of the proposed computational framework in section \ref{S:R}. In the conclusions (section \ref{S:C}), we discuss further improvements and potential future work. Meanwhile, a grid sensitivity analysis is presented in the appendix.

\section{Mathematical formulation}
\label{S:M}
\subsection{Governing equations}
\label{SS:M.1}

In the TFM, the continuity and equilibrium equations for the two phases are solved separately. Our formulation follows \citep{PF2011}, which is standard for two-phase flow solvers. However, we write the equations in a more compact form, which is useful for non-Brownian suspensions \cite{MB1999,DP14}. A general discussion of how such models are derived via averaging can be found in Drew and Passman's classic textbook \cite{DP99}. Introducing the particle volume fraction field $\phi(\bm{x},t)$, we write the governing equations for the two phases \cite{J97,B2005,G1994} as:
\begin{align}
\frac{\partial }{\partial t} \left( \rho_p \phi \right) + \nabla \cdot \left( \rho_p \bm{u}_p \phi \right)
&= 0,\label{eq::continuity_p}\\
\frac{\partial }{\partial t} \left[ \rho_f \left(1-\phi \right)  \right] + \nabla \cdot \left[ \rho_f \bm{u}_f \left(1-\phi \right) \right] &= 0,\label{eq::continuity_f}\allowdisplaybreaks\\
\frac{\partial }{\partial t} \left( \rho_p \phi \bm{u}_p \right) + \nabla \cdot \left( \rho_p  \phi \bm{u}_p \otimes \bm{u}_p \right)
&=  \nabla \cdot \bm{\Sigma}_p  + \phi \rho_p \bm{g} + \bm{f}_d, \label{eq::equilibrium_p}\\
\frac{\partial }{\partial t} \left[ \rho_f \left(1-\phi \right) \bm{u}_f \right] + \nabla \cdot \left[ \rho_f  \left(1-\phi \right) \bm{u}_f \otimes \bm{u}_f \right] &= - \ \nabla \cdot \left( p\bm{I} -  \bm{\tau}_f \right) - \bm{f}_d + (1-\phi) \rho_f \bm{g},\label{eq::equilibrium_f}
\end{align}
where the subscript `$p$' refers to a particle-phase quantity, and the subscript `$f$' refers to a fluid-phase quantity; $\otimes$ denotes the (direct) dyadic product. Also, $\rho_p$ and $\rho_f$ are the particle and fluid densities (assumed constant), $\bm{u}_p$ and $\bm{u}_f$  are the particle and fluid velocity fields, $\bm{\Sigma}_p$ is the particle-phase stress tensor, $\bm{\tau}_f$ is the deviatoric stress tensor of the fluid phase, $\bm{f}_d$ is the interphase force, and $\bm{g}$ is the gravitational acceleration vector. Here, $p=p_f+p_p$ is the `shared' pressure, which satisfies the Poisson equation in the case of an incompressible suspension, and it is given by the sum of the fluid pressure $p_f$ and the particle pressure $p_p$. Notice that, in this approach, we do \emph{not} need to attempt to calculate or introduce a phenomenological model for the  quantity $\bm{J}_\mathrm{diff}$, since we are directly computing the particle phase's velocity field $\bm{u}_p$.

\subsection{Interphase momentum transfer}

We write the interphase force $\bm{f}_d$ as the combination of a term due to the local distortion of the flow field and a generalized buoyancy. In this form, the interphase force is often referred to as the Clift drag \cite{RF1995}:
\begin{equation}
    \label{eq::fd}
    \bm{f}_d = \underbrace{K_d  \left( \bm{u}_p - \bm{u}_f \right)}_{\text{local distortion}} + \underbrace{\phi \nabla \cdot \left( \bm{\tau}_f - p\bm{I} \right) - \left(1-\phi\right) \boldsymbol{\nabla} p_p}_{\text{generalized buoyancy}} .
\end{equation}
Notice that the last term in equation \eqref{eq::fd} is a consequence of the ``shared pressure formulation,'' where the fluid phase pressure $p_f$ is eliminated using $p_f=p-p_p$. In this work, we express the drag coefficient $K_d$ as a function of the sedimentation hindrance function $f(\phi)$, which corrects the Stokes sedimentation velocity to account for the presence of neighbouring particles:
\begin{equation}
    \label{eq::Kd}
    K_d = \frac{ 9 \mu_f \phi f(\phi)^{-1} }{2 d_p^2}.
\end{equation}
Example closure expression for $f(\phi)$ will be discussed below [see equations~\eqref{eq::MillerMorris} and \eqref{eq::MorrisBoulay}].

\subsection{Rheology of the suspending fluid}
\label{SS:M.2}

We assume the suspending fluid is Newtonian, with a deviatoric stress tensor $\bm{\tau}_f$ given by
\begin{equation}
\label{eq::Sigma_f}
\bm{\tau}_f = 2\mu_f \dot{\bm{S}}_f,
\end{equation}
where as before $\mu_f$ is the dynamic fluid viscosity (assumed constant), and $\dot{\bm{S}}_f$ is the shear-rate-of-strain tensor of the fluid phase. For either phase, $\bm{S}$ is defined based on a velocity $\bm{u}$ as
\begin{equation}
\label{eq::Sdot_f}
\dot{\bm{S}} = \frac{1}{2}\left[\nabla \bm{u} + \left( \nabla \bm{u}\right)^T \right] -  \left(\nabla \cdot \bm{u} \right) \bm{I}.
\end{equation}

\subsection{Rheology of the suspended phase}
The suspended phase's rheology is given by a generalization of the expression in equation \eqref{eq::Sigma_f}:
\begin{equation}
\label{eq::Sigma_p}
\bm{\Sigma}_p = 2\mu_p \bm{\dot{S}}_p + \lambda_p\left(\nabla \cdot \bm{u}_p \right) \bm{I} + \bm{\Sigma}_s.
\end{equation}
Here, $\mu_p$ and $\lambda_p$ are the shear and bulk viscosities of the particle phase, respectively. In addition to the classic representation of the stress tensor for a compressible Newtonian fluid, equation \eqref{eq::Sigma_p} contains the extra contribution $\bm{\Sigma}_s$, which represents the anisotropic stress due to shearing of the solid phase.

The shear viscosity of the particle phase can be further decomposed as
\begin{equation}
\label{eq::nu_p}
\mu_p = \mu_{p,\text{kin}} + \mu_{p,\text{fric}},
\end{equation}
where $\mu_{p,\text{kin}}$ is a kinetic shear viscosity due to random particle kinetic (binary and instantaneous) collisions, $\mu_{p,\text{fric}}$ is a frictional shear viscosity which represents the momentum transfer due to multiple particle contact and shearing.  Similarly, we split the particle phase pressure into a kinetic particle pressure $p_{p,\text{kin}}$, arising from the Brownian motion of particles and a frictional component $p_{p,\text{fric}}$, which accounts for force chains emerging at large values of the particle volume fraction:
\begin{equation}
    \label{eq::p_p}
    p_p = p_{p,\text{kin}} + p_{p,\text{fric}}.
\end{equation}    
In dense suspensions, $\mu_{p,\text{kin}}$, $\lambda_p$ and $p_{p,\text{kin}}$ are computed, following \citep{G1994}, as functions of the granular temperature $\Theta$ \citep{G2008}, which satisfies the transport equation:
\begin{equation}
\frac{3}{2} \left[ \frac{\partial}{\partial t} \left( \rho_p \phi \Theta \right) + \nabla \cdot \left( \rho_p \phi \bm{u}_p \Theta \right)  \right]  =  \left( \bm{\Sigma}_p - p_p \bm{I}  \right):\nabla \bm{u}_p   + \nabla \cdot \left(\kappa_p \nabla \Theta \right) - \dot{q}_c + \dot{q}_{\mu} + \dot{q}_{s},
\label{eq::Theta}
\end{equation} 
where $\kappa_p$ is the granular conductivity and $\dot{q}_c$, $\dot{q}_{\mu}$ and $\dot{q}_{s}$ are source terms due to collisions, viscous dissipation and slip velocity. The expressions for these terms employed in OpenFOAM{\textsuperscript\textregistered} can be found in \cite{LS1986}.


Notice that in the parameter regimes in which particle migration is relevant, kinetic collisions are negligible and so  the transport equation for $\Theta$ plays no role in this limit, therefore it can be neglected, as is generally done in works on non-Brownian suspensions. However, in our OpenFOAM{\textsuperscript\textregistered} implementation of the TFM that follows, we retain equation~\eqref{eq::Theta} and all the related terms. Doing so allows us (in the future) to employ the proposed TFM to simulate transitional suspension dynamics, i.e., from dense (non-kinetic) to dispersed (kinetic) particulate phases. However, such complex flow regimes will not be explored in this work; we focus only on non-kinetic flows in which the granular temperature $\Theta$ can be considered constant.

The anisotropic shear-induced stress tensor $\bm{\Sigma}_s$ is generally represented by means of an anisotropy tensor $\bm{Q}$ \cite{MB1999}, specifically:
\begin{equation}
    \label{eq::sigma_s}
    \bm{\Sigma}_s = - \mu_f \eta_{\text{N}} (\phi) \dot{\gamma}_{\text{eff}} \bm{Q}.
\end{equation}
Here, $\eta_{\text{N}}$ is the normal scaled viscosity, and $\dot{\gamma}_{\text{eff}}$ is the effective shear rate defined as
\begin{equation}
    \label{eq::gamma_ff}
    \dot{\gamma}_{\text{eff}} = \sqrt{2 \dot{\bm{S}}:\dot{\bm{S}}} + \dot{\gamma}_{\text{NL}}.
\end{equation}
Here, $\dot{\gamma}_{\text{NL}}$ is the non-local shear rate, which is often employed to ensure $\dot{\gamma}_{\text{eff}} \neq 0$, for example, at the centerline of a channel. The addition of $\dot{\gamma}_{\text{NL}}$ in equation~\eqref{eq::gamma_ff} is seen as a way to overcome the breakdown of continuum models when describing phenomena occurring at the particle scale \cite{MB98,NB94,MB1999,MS95}. When such a breakdown occurs, the non-local contribution takes into account the effect of the average stress on a scale on the order of the particle diameter. 

An exhaustive description of this phenomenon can be found in \cite{MM2006}, where the following expression for $\dot{\gamma}_{\text{NL}}$, valid for channel flows, is proposed:
\begin{equation}
    \label{eq::gammadotNL}
    \dot{\gamma}_{\text{NL}} = a_s \frac{u_p^{\text{max}}}{L_{\text{ch}}}.
\end{equation}
Here, $L_{\text{ch}}$ is the characteristic length of the channel, $a_s$ is a model constant, and $u_p^{\text{max}}$ is the maximum value of the particle velocity in the channel. Another expression, which does not depend on the flow conditions, was proposed by Gao et al.~\cite{GXG09} by fitting a large amount of experimental data:
\begin{equation}
    \dot{\gamma}_{\text{NL}} = 0.0176\left( \phi_c \right)^{-2.91},
\end{equation}
where $\phi_c$ is the concentration at the center of the channel, which is  unknown \textit{a priori}.

The anisotropy tensor $\bm{Q}$ is represented in the classic Cartesian tensor form $\bm{Q}=Q^{ij}\bm{e}_i \otimes \bm{e}_j$ (Einstein summation notation implied) and diagonalized by employing a velocity-field-based coordinate system with orthonormal axes:
\begin{equation}
    \label{eq::Q}
    \bm{Q} = \sum_{i=1}^3 \lambda_i\left( \phi \right) \bm{e}_i \otimes \bm{e}_i.
\end{equation}
Here, $\lambda_i \left( \phi \right)$ are the anisotropy weight functions, and $\bm{e}_i$ are the unit vectors in the direction of the flow ($i=1$), gradient ($i=2$) and vorticity ($i=3$) of the particle phase velocity. However, while the above definition of the unit vectors is straightforward in unidirectional flows, $\bm{e}_2$ cannot be straightforwardly calculated in general curvilinear three-dimensional flows. Therefore, in this work, we define the unit vectors $\{\bm{e}_i\}_{i=1,2,3}$ as follows:
\begin{equation}
    \label{eq::unitVectors_Q}
    \bm{e}_1 = \frac{\bm{u}_p}{|\bm{u}_p|}, \qquad \bm{e}_3 = \frac{\nabla \times \bm{u}_p}{|\nabla \times \bm{u}_p|}, \qquad \bm{e}_2 = \bm{e}_1 \times \bm{e}_3,
\end{equation}
where $\times$ denotes the vector (cross) product. Notice that, by employing equation \eqref{eq::unitVectors_Q}, we calculate an ``implicit'' gradient direction using the properties of the vector product. Specifically, $\bm{e}_2$ is simply defined as being normal to both the particle velocity field and its curl.

\subsection{Closure models for non-Brownian suspensions}

In the present work, we incorporate closure models that have been developed for mixture theories (like the DFM or the ``standard'' form of the SBM currently employed in applications) into a TFM framework. As discussed in \cite{GP18}, such models describe the overall suspension viscosity and pressure, therefore they do not differentiate between particle-particle lubrication mediated interaction and frictional contacts. Consequently, such models directly provide a closure for $\mu_{p,\text{fric}}$ from equation~\eqref{eq::nu_p}, without addressing the fine details of the suspension's microstructure during shear-thickening, especially as it approaches jamming \cite{Morris2018}. On the other hand, the effect of the frictional pressure $p_{p,\text{fric}}$ from equation~\eqref{eq::p_p} will be absorbed in the anisotropic stress $\bm{\Sigma}_s$. A similar approach was employed in \cite{DLVS2018}, where the frictional pressure and viscosity where not included in the governing equations.

Thus, we employ a general expression similar to that proposed in \cite{MB1999} to close the particle phase viscosity:
\begin{equation}
    \label{eq::closure_mup}
    \frac{\mu_{p,\text{fric}}}{\mu_f} = a_{\mu} + b_{\mu} \phi \left( 1 - \frac{\phi}{\phi_m} \right)^{-1} + c_{\mu}  \left( 1 - \frac{\phi}{\phi_m} \right)^{-2},
\end{equation}
which returns the closure from \cite{MB1999} for $a_{\mu}=0$, $b_{\mu}=2.5$ and $c_{\mu}=0.1$, and the closure from \cite{MP56} for  $a_{\mu}=-1$, $b_{\mu}=0$ and $c_{\mu}=1$. These models have been shown to give the best agreement with experiments when employed within the framework of the SBM \cite{DLLM2013}. In equation~\eqref{eq::closure_mup}, $\phi_m$ is the maximum allowed particle volume fraction.

For the normal scaled viscosity $\eta_{\text{N}}$, we employ the expression proposed in \cite{MB1999}:
\begin{equation}
    \label{eq::etaN}
    \eta_{\text{N}} (\phi)= K_{\text{N}}  \left( \frac{\phi}{\phi_m} \right)^2  \left( 1 - \frac{\phi}{\phi_m} \right)^{-2},
\end{equation}
where $K_{\text{N}}$ is usually set to $0.75$. Equation \eqref{eq::etaN} also returns the model proposed in \cite{DLL13} for $K_{\text{N}}=1.08$, and the one proposed in \cite{BGP2011} for $K_{\text{N}}=1$.

The sedimentation hindrance function $f(\phi)$ is modeled using the expression provided in \cite{MM2006}:
\begin{equation}
    \label{eq::MillerMorris}
    f(\phi) = \left(1 - \frac{\phi}{\phi_m} \right) \left(1 - \phi \right)^{\alpha-1}, \quad \alpha \in \left[ 2,5 \right].
\end{equation}
This expression was chosen to ensure that particle migration becomes weaker as $\phi\to\phi_m$. Another expression often employed can be found in \cite{MB1999}:
\begin{equation}
    \label{eq::MorrisBoulay}
        f(\phi) =\left(1 - \phi \right)^{\alpha}, \quad \alpha \in \left[ 2,5 \right].
\end{equation}

Finally, several expressions for $\lambda_i$ have been proposed in the literature \cite{DLL13, BGP2011,MM2006}. Taking each to be a particular constant is a common choice, but \cite{DLL13} also proposed volume-fraction-dependent expressions:
\begin{equation}
\label{eq::lambda_DLL}
    \lambda_1(\phi) = 1, \qquad \lambda_2(\phi) = 0.81 \frac{\phi}{\phi_m} +0.66, \qquad \lambda_3(\phi) = -0.0088 \frac{\phi}{\phi_m} + 0.54.
\end{equation}

\section{Numerical formulation}
\label{S:N}

The governing equations \eqref{eq::continuity_p}--\eqref{eq::equilibrium_f} are solved in a coupled manner using a modified version of the \emph{twoPhaseEulerFoam} solver \cite{PF2011} in OpenFOAM{\textsuperscript\textregistered}  \cite{WTJF98}, an open-source library designed for implementing finite volume methods \cite{MMD2016}. Momentum predictors are obtained employing the partial elimination algorithm  \cite{PF2011}, which allow us to decouple the phase momentum equations. In this work, we extend \emph{twoPhaseEulerFoam} to include models for shear-induced migration and to employ the anisotropic stress tensor in place of the frictional pressure. 
In fact, the main idea behind the approach used in OpenFOAM\textsuperscript\textregistered\ is to include the effect of the particle pressure in the suspended (dispersed) phase's continuity equation in an implicit manner.

\subsection{Discretized momentum equations}

In order to clearly illustrate our modifications to the original algorithm described in \cite{PF2011}, below we consider an incompressible suspension in which the phases have constant and equal density (therefore, we drop the gravitational force). Under these assumptions, we can write the semi-discrete momentum equations as
\begin{align}
    \label{eq::momentum_sd_p}
    \mathbb{A}_p \bm{u}_p &= \mathbb{H}_p - \mu_f\nabla \cdot \left( \eta_{\text{N}} \dot{\gamma}_{\text{eff}} \bm{Q}  \right) - \phi\nabla p + K_d \left( \bm{u}_p - \bm{u}_f \right),\\
    \label{eq::momentum_sd_f}
    \mathbb{A}_f \bm{u}_f &= \mathbb{H}_f - \left(1-\phi \right) \nabla p + K_d \left( \bm{u}_p + \bm{u}_f \right),
\end{align}
where the matrices $\mathbb{A}_p$ and $\mathbb{A}_f$ are the diagonals of the matrices $\mathbb{M}_f$ and $\mathbb{M}_p$ arising from the discretization of the respective momentum equations, with the exception of the undiscretized terms retained in equations \eqref{eq::momentum_sd_p} and \eqref{eq::momentum_sd_f}. Meanwhile, the vectors $\mathbb{H}_p$ and $\mathbb{H}_f$ are given by:
\begin{equation}
    \label{eq::H}
        \mathbb{H}_p = \left( \mathbb{A}_p - \mathbb{M}_p  \right)\bm{u}_p + \mathbb{Q}_p, \qquad
        \mathbb{H}_f = \left(\mathbb{A}_f - \mathbb{M}_f   \right)\bm{u}_f + \mathbb{Q}_f,
\end{equation}
where $\mathbb{Q}_p$ and $\mathbb{Q}_f$ are the source terms arising from the discretization of the momentum equations (volumetric sources, face-tangential corrections, etc.).

The key idea of the algorithm is to split the anisotropic stress tensor flux in two contributions: one due to the flux arising from a gradient in the particle volume fraction and one due to the flux arising from a gradient in the shear rate or the anisotropy tensor. Specifically,
\begin{equation}
    \label{eq::splitting}
         \nabla \cdot \left( \eta_{\text{N}} \dot{\gamma}_{\text{eff}} \bm{Q}  \right) = \dot{\gamma}_{\text{eff}} \left( \frac{d \eta_{\text{N}}}{d \phi} \right) \nabla \phi \cdot  \bm{Q} + \eta_{\text{N}} \nabla \cdot \left( \dot{\gamma}_{\text{eff}} \bm{Q} \right).
\end{equation}
This idea is similar to the diffusive flux model \cite{vsb10,pabga92}, wherein the forcing terms due to the shear-induced migration are written as the gradient of a chemical potential. However, in the present model, we employ a tensor potential instead of a scalar potential.
The rightmost term in the decomposition in equation \eqref{eq::splitting} is then subtracted from the vector $\mathbb{H}$, so that equation \eqref{eq::momentum_sd_p} becomes:
\begin{equation}
    \label{eq::momentum_Hstar}
     \mathbb{A}_p \bm{u}_p = \mathbb{H}^{*}_p - \dot{\gamma}_{\text{eff}} \left( \frac{d \eta_{\text{N}}}{d \phi} \right) \nabla \phi \cdot  \bm{Q}, \qquad \mathbb{H}^{*}_p = \mathbb{H}_p -  \eta_{\text{N}} \nabla \cdot \left( \dot{\gamma}_{\text{eff}} \bm{Q} \right).
\end{equation}

\subsection{Pressure equation}

The incompressibility condition on the suspension requires surface integrals of the volumetric flux to be zero in every cell. In other words, the total volumetric flux $\varphi$ must be such that the mixture velocity field $\bm{u}_\mathrm{mix}$ defined in equation~\eqref{eq::suspension_U} is divergence free:
\begin{equation}
   \label{eq::div_free}
    \nabla \cdot \bm{u}_\mathrm{mix} = \nabla \cdot \left( \phi \bm{u}_p \right) + \nabla \cdot \left((1- \phi) \bm{u}_f \right) = 0,
\end{equation}
In the finite volume method,  equation~\eqref{eq::div_free} is satisfied if the sum of the phase volumetric fluxes vanishes in each cell $c$, i.e.,
\begin{equation}
   \label{eq::flux_free}
    \int_{\mathcal{V}_c} \nabla \cdot \left[ \phi \bm{u}_p  + (1- \phi) \bm{u}_f  \right] d\mathcal{V}_c = 
    \sum_{N_{\mathrm{cf}}} \oint_{S_{\mathrm{cf}}} \left[ \phi \bm{u}_p  + (1- \phi) \bm{u}_f  \right] \cdot d\bm{A}_{\mathrm{cf}}
    = \sum_{N_{\mathrm{cf}}} \left[ \phi_{\mathrm{cf}} \varphi_p + (1-\phi)_{\mathrm{cf}} \varphi_f  \right],
\end{equation}
where $\mathcal{V}_c$ is the volume of cell $c$, $\bm{A}_{\mathrm{cf}}$ is the area normal to face $\text{cf}$, $N_{\text{cf}}$ is the number of faces of cell $c$, and $\phi_{\mathrm{cf}}$ is the particle volume fraction interpolated at face $\mathrm{cf}$. Additionally, in equation \eqref{eq::flux_free}, we introduced the volumetric fluxes of the particle phase $\varphi_p$ and the fluid phase $\varphi_f$:
\begin{equation}
    \label{eq::fluxes}
    \varphi_p = \oint_{S_{\mathrm{cf}}} \bm{u}_p \cdot d\bm{A}_{\mathrm{cf}}, \qquad \varphi_f = \oint_{S_\mathrm{{cf}}} \bm{u}_f \cdot d\bm{A}_{\mathrm{cf}}.
\end{equation}
Notice that the phase fluxes are `scalar fields' defined at cell faces rather than at cell centers.

Next, decoupled equations for $\bm{u}_p$ are obtained by substituting equation \eqref{eq::momentum_sd_f} into equation \eqref{eq::equilibrium_p}. The same approach is employed to obtain decoupled equations for $\bm{u}_f$. Therefore, the volumetric phase fluxes can be expressed as \citep{PF2011}:
\begin{align}
    \label{eq::flux_p}
    \varphi_p &= \frac{\bm{A}_{\mathrm{cf}}}{\zeta_p} \cdot \left[ \left( \mathbb{H}^{*}_p +  K_d \beta_f \mathbb{H}_f \right)_{\mathrm{cf}}  - \left( \phi + K_d \beta_f (1-\phi)  \right)_{\mathrm{cf}} \left(\nabla p \right)_{\mathrm{cf}} -\left( \dot{\gamma}_{\text{eff}} \frac{d \eta_{\text{N}}}{d \phi}  \nabla \phi \cdot  \bm{Q}  \right)_{\mathrm{cf}}    \right],\\
    \label{eq::flux_f}
    \varphi_f &= \frac{\bm{A}_{\mathrm{cf}}}{\zeta_f} \cdot \left[ \left( \mathbb{H}^{*}_f +  K_d \beta_p \mathbb{H}_p \right)_{\mathrm{cf}}  - \left( (1-\phi) + K_d \beta_p \phi  \right)_{\mathrm{cf}} \left(\nabla p \right)_{\mathrm{cf}} + \left( K_d \beta_p \dot{\gamma}_{\text{eff}} \frac{d \eta_{\text{N}}}{d \phi}  \nabla \phi \cdot  \bm{Q}  \right)_{\mathrm{cf}}    \right],
\end{align}
where the subscript `cf' indicates interpolation at cell faces, and
\begin{equation}
    \label{eq::betas}
    \beta_p = \frac{1}{\mathbb{A}_p + K_d}, \qquad \beta_f = \frac{1}{\mathbb{A}_f + K_d}, \qquad
   \zeta_p = \mathbb{A}_p - \beta_p K^2_d + K_d, \qquad \zeta_f = \mathbb{A}_f - \beta_f K^2_d + K_d.
\end{equation}
Notice that unlike \cite{PF2011}, our flux equations \eqref{eq::flux_p} and \eqref{eq::flux_f} do {not} have a term of the kind $\bm{A}_{\mathrm{cf}} \cdot \left( \nabla \phi\right)_{\mathrm{cf}}$ because we are employing an \emph{anisotropic} stress tensor rather than an isotropic particle pressure. At this stage, this distinction does not constitute a significant computational difference since the term is not updated within the pressure corrector.

The pressure equation is obtained by substituting equations \eqref{eq::flux_p} and \eqref{eq::flux_f} into equation \eqref{eq::div_free}. After solving for $p$, the new pressure field is used in equations \eqref{eq::flux_p} and \eqref{eq::flux_f} to update the phase fluxes.

\subsection{Continuity equation for the suspended phase}
\label{subsec::conteq}

The numerical approach employed has a major impact on the overall algorithm's stability, especially when the particle volume fraction approaches the close-packing limit. To this end, consider the semi-discretized continuity equation for the suspended (dispersed) phase:
\begin{equation}
    \label{eq::continuity_p_flux}
    \frac{\partial \phi}{\partial t} + \nabla \cdot \left( \phi_{\mathrm{cf}} \varphi_s^* \right) - \nabla \cdot \left[ \frac{\phi_{\mathrm{cf}}\bm{A}_{\mathrm{cf}}}{\zeta_p} \cdot \left( \dot{\gamma}_{\text{eff}} \frac{d \eta_{\text{N}}}{d \phi}  \nabla \phi \cdot  \bm{Q}  \right)_{\mathrm{cf}}\right] = 0,
\end{equation}
where $\varphi_s^*$ is the volumetric flux of the particle phase $\varphi_p$ from equation~\eqref{eq::flux_p} with the contribution from the anisotropic tensor split from it. Notice that the divergence in equation~\eqref{eq::continuity_p_flux} is a semi-discretized operator that acts on face-interpolated scalar fields, and it should be interpreted as a sum over cell faces following a domain discretization. Since  $d \eta_{\text{N}}/d \phi$ increases dramatically close to the maximum packing fraction \cite{GP18}, in the standard \emph{twoPhaseEulerFoam} the equivalent rightmost term on the left-hand side of equation \eqref{eq::continuity_p_flux} is discretized implicitly to avoid instabilities in this regime. However, in the present context, the term cannot be immediately discretized as is because it does not involve just $\bm{A}_{\mathrm{cf}} \cdot (\nabla \phi )_{\mathrm{cf}}$, which can be discretized using standard divergence schemes like upwind, but it involves the double scalar product $\bm{A}_{\mathrm{cf}} \cdot \left(\nabla \phi \cdot  \bm{Q}  \right)_{\mathrm{cf}}$ instead. Thus, this term requires more than just discretizing the face normal gradient.

To overcome this difficulty, and extract new terms from equation~\eqref{eq::continuity_p_flux} that can be discretized implicitly in time, we decompose the anisotropy tensor into its hydrostatic and deviatoric components:
\begin{equation}
    \label{eq::decompose_Q}
    \bm{Q} = \left( \tr \bm{Q}\right) \bm{I} + \bm{Q}_{\mathrm{dev}}, \qquad \bm{Q}_{\mathrm{dev}} \equiv \bm{Q} - \left( \tr \bm{Q} \right)  \bm{I},
\end{equation}
where $\tr \bm{Q} = \sum_{i=1}^3 \bm{Q}_{ii}$ is the trace of $\bm{Q}$. 
Therefore, equation \eqref{eq::continuity_p_flux} can be rewritten (still considering our semi-discretized definition of the divergence operator) as:
\begin{equation}
    \label{eq::continuity_p_flux2}
    \frac{\partial \phi}{\partial t} + \nabla \cdot \left( \phi_{\mathrm{cf}} \varphi_s^* \right) - \nabla \cdot \left[  \left( \dot{\gamma}_{\text{eff}} \frac{d \eta_{\text{N}}}{d \phi } \tr \bm{Q}  \right)_{\mathrm{cf}} \frac{\bm{A}_{\mathrm{cf}}}{\zeta_p} \cdot \left( \nabla \phi \right)_{\mathrm{cf}}\right] = \nabla \cdot \left[ \frac{\phi_{\mathrm{cf}}\bm{A}_{\mathrm{cf}}}{\zeta_p} \cdot \left( \dot{\gamma}_{\text{eff}} \frac{d \eta_{\text{N}}}{d \phi}  \nabla \phi \cdot  \bm{Q}_{\mathrm{dev}}  \right)_{\mathrm{cf}}\right].
\end{equation}
Notice that now only the term on the right-hand side of equation~\eqref{eq::continuity_p_flux2} cannot be discretized implicitly.
The approach used in OpenFOAM{\textsuperscript\textregistered} for solving equation \eqref{eq::continuity_p_flux2} consists of solving the advection equation $ {\partial \phi}/{\partial t} + \nabla \cdot \left( \phi_{\mathrm{cf}} \varphi_s^* \right) = 0$ explicitly using the MULES (Multidmensional Universal Limiter for Explicit Solutions) scheme, which is based on the Flux-Corrected Transport (FCT) framework \cite{Z1979,BB1997}. The remaining terms are solved implicitly in time. External iterations with relaxation are still required to update the last term in equation~\eqref{eq::continuity_p_flux2} and obtain an accurate and stable solution. However, their number and the value of the relaxation coefficient are now controlled only by the deviatoric part of $\bm{Q}$. 

As a remark, it should be pointed out that equation \eqref{eq::continuity_p_flux2} is parabolic, while the original continuity equation was hyperbolic. This change of type may be an issue, especially regarding the choice of appropriate boundary conditions for $\phi$. Therefore, in order to preserve the hyperbolicity of the continuity equation, an implicit formulation of the anisotropic stress should be employed only when facing severe stability issues.

\subsection{Numerical solution strategy: Description of the scheme}

\begin{figure}
     \centering
     \includegraphics[width=\linewidth]{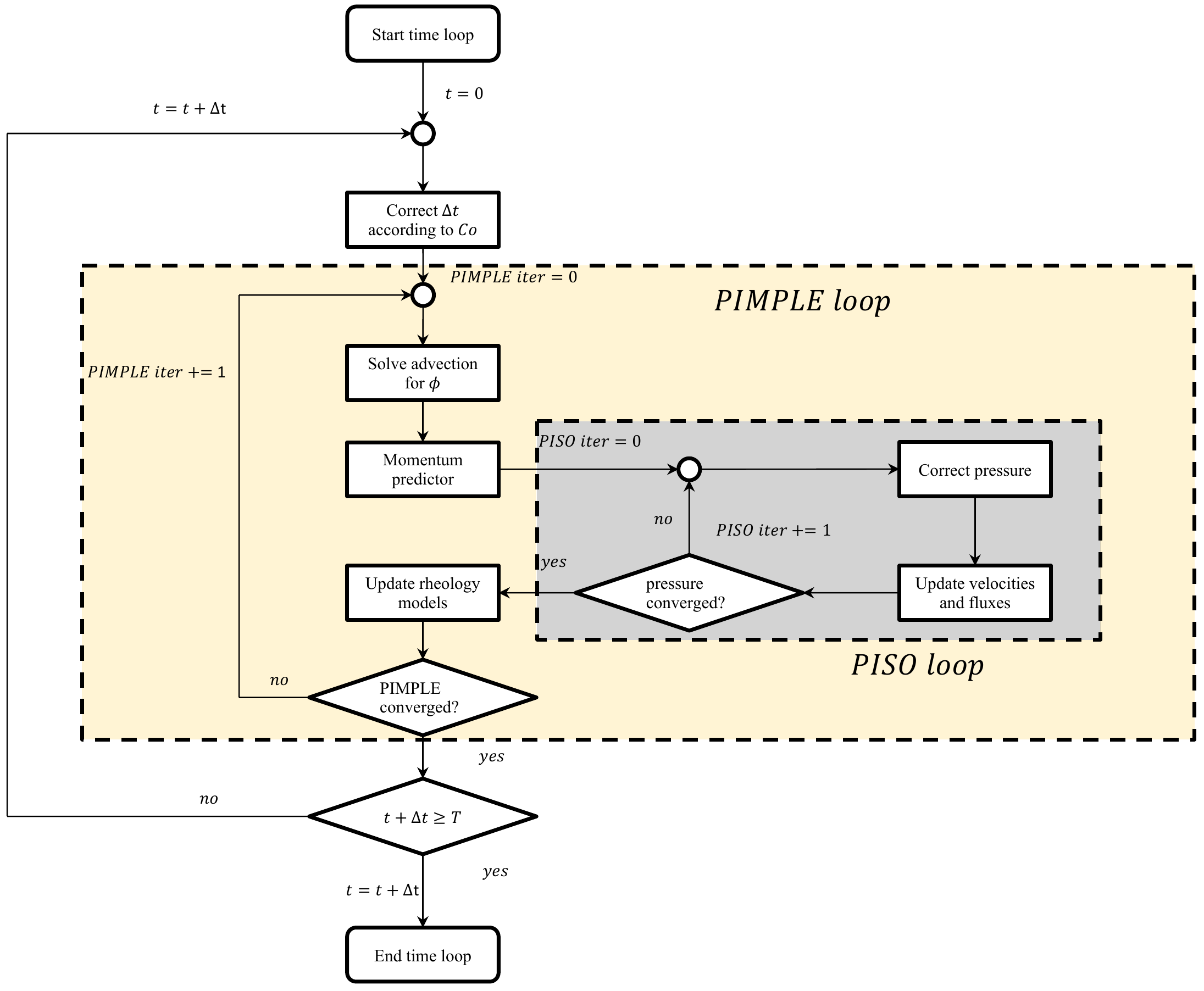}
     \caption{Flowchart illustrating the solution algorithm composed by external iterations (\textit{PIMPLE} loop) and pressure correction iterations (\textit{PISO} loop). Here $T$ is the fixed end time of the simulation. Notice that this is the same algorithm employed in $twoPhaseEulerFoam$.}
     \label{fig::loops}
\end{figure}

The solver employs a \textit{PIMPLE} algorithm, which consists of a \textit{PISO} \cite{FP_chapter} pressure corrector together with external fixed point iterations to couple the phase momentum equations and the continuity equation for the particle phase, as depicted schematically in figure~\ref{fig::loops}. The solver employs a dynamic time stepping based on the Courant number:
\begin{equation}
    \label{eq::Co}
     Co = \frac{\varphi_{Co} \Delta t}{ V }, \qquad \varphi_{Co} = \text{max}\left( \sum_{\mathrm{cf}} \varphi_f, \sum_{\mathrm{cf}} \left( \varphi_f - \varphi_p \right) \right),
\end{equation}
where $\Delta t$ is the time step and $V$ is the cell volume field. Notice that the velocity field employed in the definition of $Co$ is the maximum between the total suspension flux and the total relative flux in each cell. 

The convergence criteria for both the external \textit{PIMPLE} iterations and the \textit{PISO} correctors can be based on the residuals or the number of iterations. In this work, we always limited the number of iterations and set the minimum number of linear solver iterations to one. This choice was made to control residual oscillations and prevent the solver from leaving the loops too early.

\subsection{Boundary conditions}

Generally, no-slip and no penetration boundary conditions are employed to model walls bounding the flows of non-Brownian suspensions (see, e.g., \cite{DLL13}). The no penetration boundary condition is implemented in OpenFOAM{\textsuperscript\textregistered} as a \emph{slip} boundary condition, which is fundamentally a symmetric boundary condition that implies no flux normal to the boundary. Such a boundary condition imposes:
\begin{equation}
     \bm{J}_{\phi} \cdot \bm{n} = 0, \quad\text{with}\quad \bm{J}_{\phi} = \bm{u}_s \phi,
\end{equation}
where $\bm{n}$ is the unit normal vector to the boundary.

\section{Testing and validation}
\label{S:R}
\subsection{Planar Poiseuille flow of a suspension}
\label{SS:R.1}

\begin{figure}
    \centering
    \includegraphics[width=0.9\linewidth]{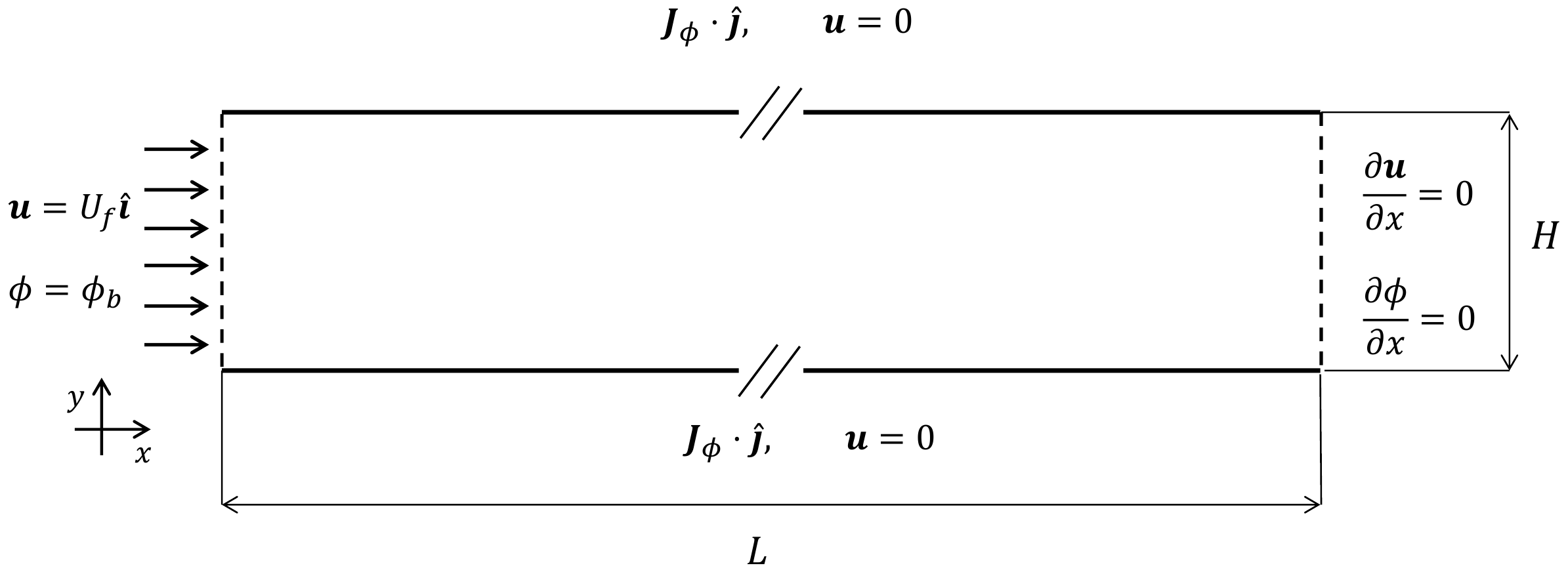}
    \caption{Schematic representation of the two-dimensional (2D) channel geometry, indicating the boundary conditions for the velocity field (both the particle and fluid velocity are represented by a single symbol $\bm{u}$) and the particle phase's volumetric concentration field $\phi$. Here, $\hat{\bm{\imath}}$ and $\hat{\bm{\jmath}}$ are the unit normal vector in the $x$-direction and $y$-direction respectively, and $U_f$ is the fluid velocity in the particle-based Reynolds number in equation~\eqref{eq::Rep_cond}.}
    \label{fig::channel_scheme}
\end{figure}

We apply the proposed TFM numerical formulation to the problem of suspension flow between two infinite parallel plates. This problem has been extensively studied in many experimental works \cite{KHL1994,LL1998,SMW2007,IBP2007,GXG09}, which consistently observed a particle flux towards the center of the channel and, thus, an inhomogeneous particle distribution. However, it was argued that the laser-Doppler velocimetry methodology employed by Lyon and Leal \cite{LL1998} significantly underestimates the particle volume concentration near the walls \cite{MB1999}. Recently, Drijer et al.~\cite{DLVS2018} performed an experiment employing a fast confocal microscope (as in \cite{SMW2007}) and were able to match the concentration profile with results from their `hybrid' SBM-TFM simulations in STAR-CCM+. We employ a conduit identical to that used by Dbouk et al.~\citep{DLLM2013}, with a ratio between channel height and particle diameter of $H/d_p = 18$. A schematic of the channel geometry and notation is shown in figure \ref{fig::channel_scheme}. 

The channel length $L$ is chosen to satisfy the condition for a fully developed concentration profile \citep{NB94}:
\begin{equation}
    \label{eq::LH_FD}
    \frac{L}{H} \geq \frac{1}{6g(\phi_b)} \left( \frac{H}{d_p} \right)^2.
\end{equation}
The function $g(\phi_b)$ represents the dependence of the shear-induced diffusion on the bulk (average) particle concentration. This function is generally taken to be (see \citep{la87}):
\begin{equation}
    \label{eq::g_phi}
    g(\phi_b) = \frac{1}{3} \phi_b^2 \left( 1 + \frac{1}{2}e^{8.8\phi_b} \right).
\end{equation}
Notice that, in fully periodic domains, the condition in equation \eqref{eq::LH_FD} gives the minimum time required to achieve a fully developed concentration profile. In order to drive the flow through the fully periodic channel, a body force is applied that ensures an average value $U_f$ of the fluid (or particle) velocity field. This value is chosen such that the resulting particle Reynolds number [defined in equation~\eqref{eq::Rep_cond}] is small, specifically $Re_p \approx 10^{-3} \ll 1$. Then, the condition in equation~\eqref{eq::LH_FD} can be re-expressed as:
\begin{equation}
    \label{eq::t_FD}
    t \geq \frac{H}{U_f}\frac{1}{6g(\phi_b)} \left( \frac{H}{d_p} \right)^2.
\end{equation}

\begin{figure}[ht!]
\centering
\begin{subfigure}{.5\textwidth}
  \centering
  \includegraphics[width=0.95\linewidth]{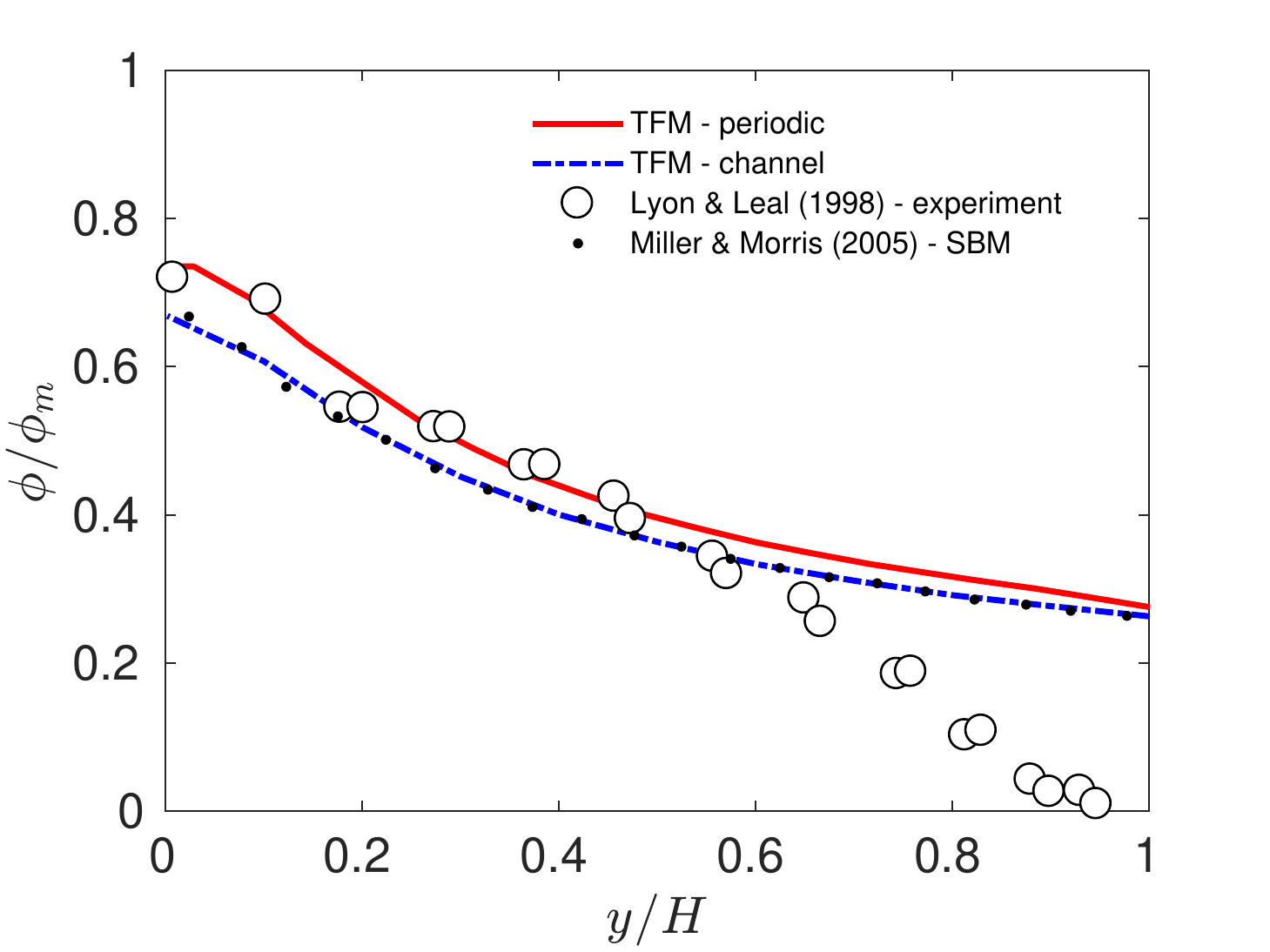}
  \caption{Cross-sectional concentration profile, $\phi_b = 0.3$}
  \label{fig:channel1}
\end{subfigure}%
\begin{subfigure}{.5\textwidth}
  \centering
  \includegraphics[width=0.95\linewidth]{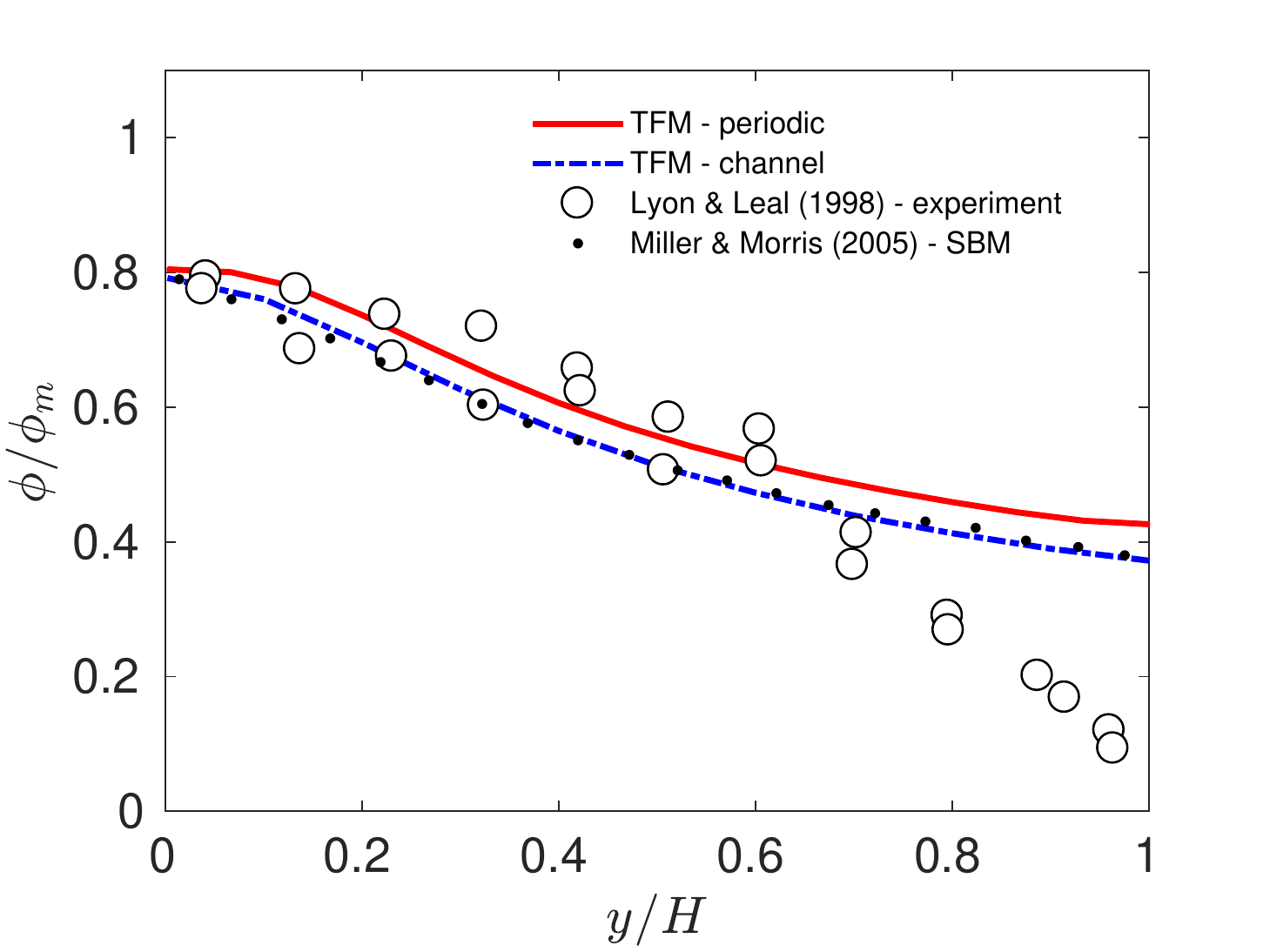}  
  \caption{Cross-sectional concentration profile, $\phi_b=0.4$}
  \label{fig:channel2}
\end{subfigure}
\begin{subfigure}{.5\textwidth}
  \centering
  \includegraphics[width=0.95\linewidth]{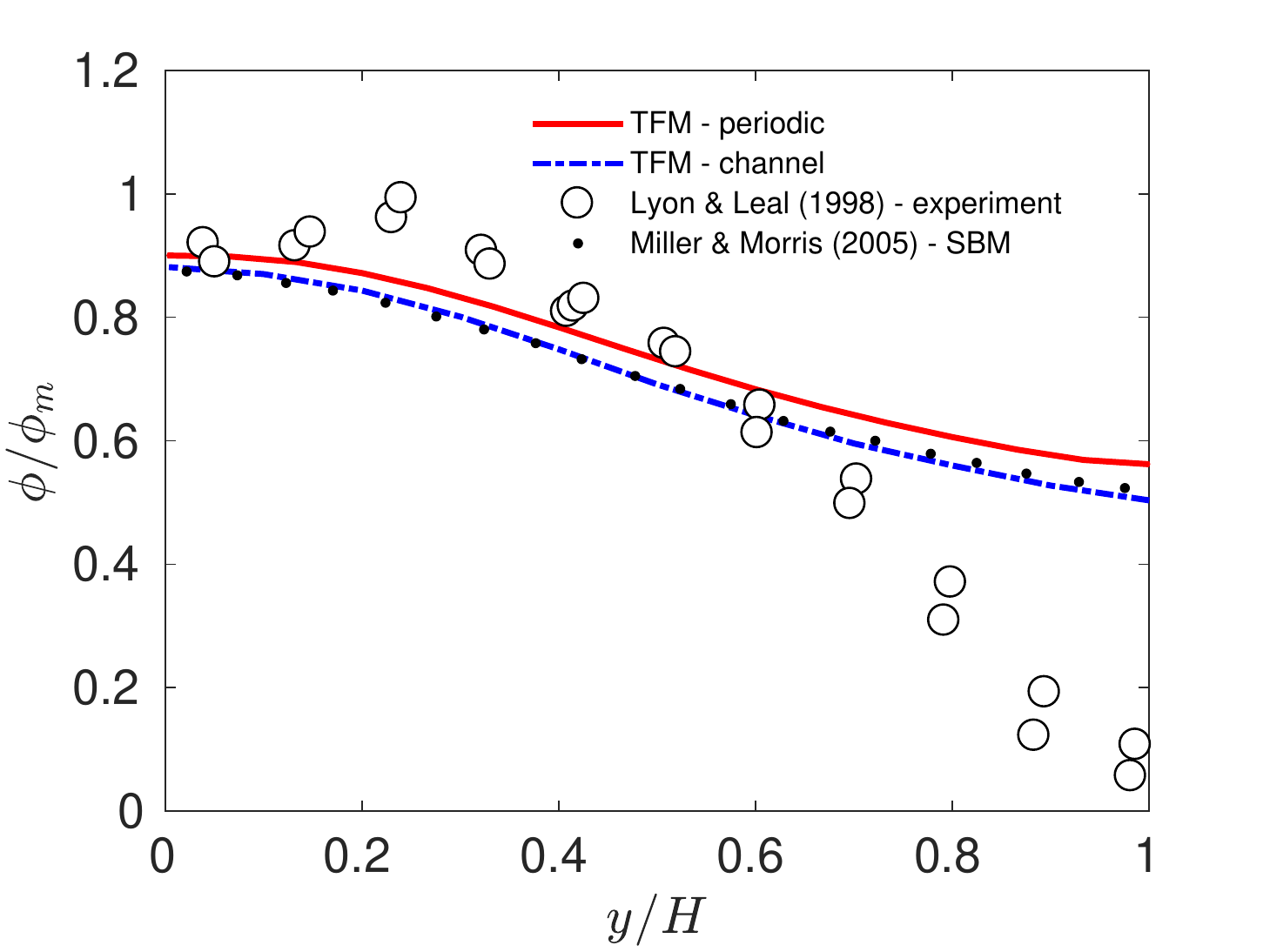}  
  \caption{Cross-sectional concentration profile, $\phi_b=0.5$}
  \label{fig:channel3}
\end{subfigure}%
\begin{subfigure}{.5\textwidth}
  \centering
  \includegraphics[width=0.95\linewidth]{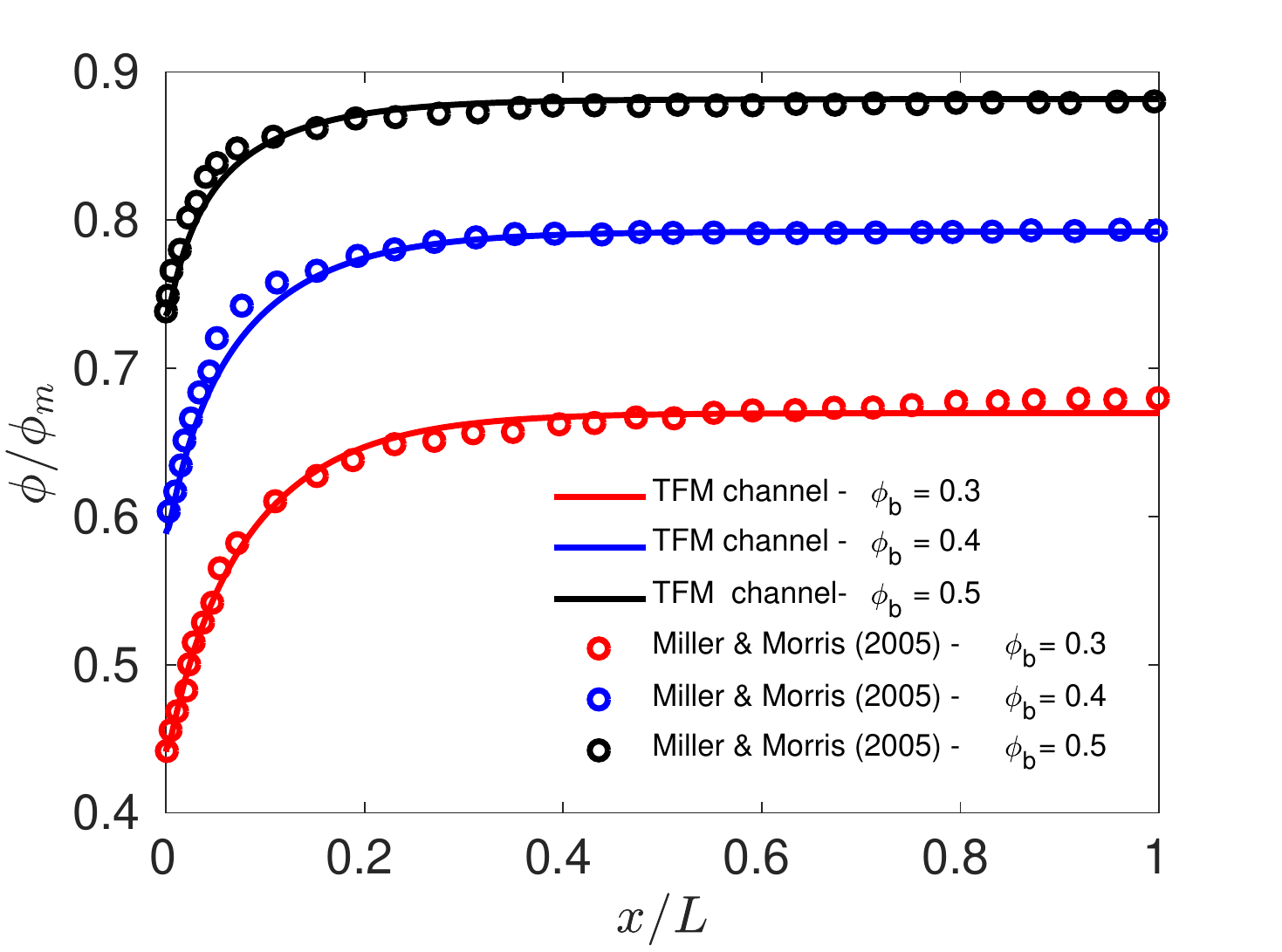}  
  \caption{Axial concentration profiles for all $\phi_b$}
  \label{fig:channel4}
\end{subfigure}
\caption{Results for the 2D pressure-driven channel flow from our TFM implementation compared against experimental results from Lyon and Leal \cite{LL1998} and the SBM results of Miller and Morris \citep{MM2006}.}
\label{fig::channel}
\end{figure}

We study particles with a diameter $d_p = 50 ~\mu \text{m}$ suspended in a fluid with $\mu_f = 0.48 ~\text{Pa s}$ and $\rho_f = \rho_p = 1.19 ~\text{g} \, \text{cm}^{-3}$. We employ the rheological closures suggested by Miller and Morris \citep{MM2006}, since those were shown to produce the most accurate results \citep{DLLM2013}. The mesh consists of $20$ cells in the direction of the gap. For the case of the full channel, the mesh in the flow direction is made of $100$ cells with a growth factor of $50$ as in \citep{DLLM2013}. Closures adopted are detailed in table \ref{tab::closuresPoiseuille}. As shown in figure \ref{fig::channel}, our two-fluid model (TFM) is able to accurately reproduce results from the suspension balance model (SBM) under its customary simplifications.

\begin{table}
    \centering
    \begin{tabular}{|c|c|l|}
    \hline
      closure  & expression & coefficients  \\
      \hline
        $f(\phi)$ & equation \eqref{eq::MillerMorris} & $\alpha=4$ \\
        $\eta_{\text{N}} (\phi)$ & equation \eqref{eq::etaN} & $K_{\text{N}}=0.75$ \\
        $\mu_p/\mu_f$ & equation \eqref{eq::closure_mup} & $a_{\mu} = -1$, $b_{\mu} = 0$, $c_{\mu} = 1$ \\
        $\lambda_i(\phi)$ & constant values & $\left\lbrace 1.0, \; 0.8, \; 0.5 \right\rbrace$ \\
        $\phi_m$ & constant value & $0.68$ \\
        $\dot{\gamma}_{\text{NL}}$ & equation \eqref{eq::gammadotNL} & $a_s = d_p/H$ \\
    \hline
    \end{tabular}
    \caption{Closure models and parameters used for the parallel plates (planar Poiseuille) configuration.}
    \label{tab::closuresPoiseuille}
\end{table}

Unlike previous computational studies \cite{YM13,DLL13,SMW2007}, we compare the results obtained with a periodic 2D domain to those obtained with an entire channel domain. Figure \ref{fig::channel} shows that a difference exists between the two configurations, which can be attributed to the inlet boundary condition. In each situation, the domain is initialized with a fixed average particle volume concentration $\phi_b$, which is conserved in the case of a periodic channel. In the full channel, $\phi_b$ likewise correspond to the fixed particle concentration at the inlet. The latter value is not conserved along the channel. Instead, for each axial cross-section $A$, the average volumetric flux $\langle\bm{u}_p\phi\rangle_A$ is conserved. This leads to an overall over-prediction of the particle concentration in a fully periodic configuration.

\begin{figure}
\centering
  \includegraphics[width=\linewidth]{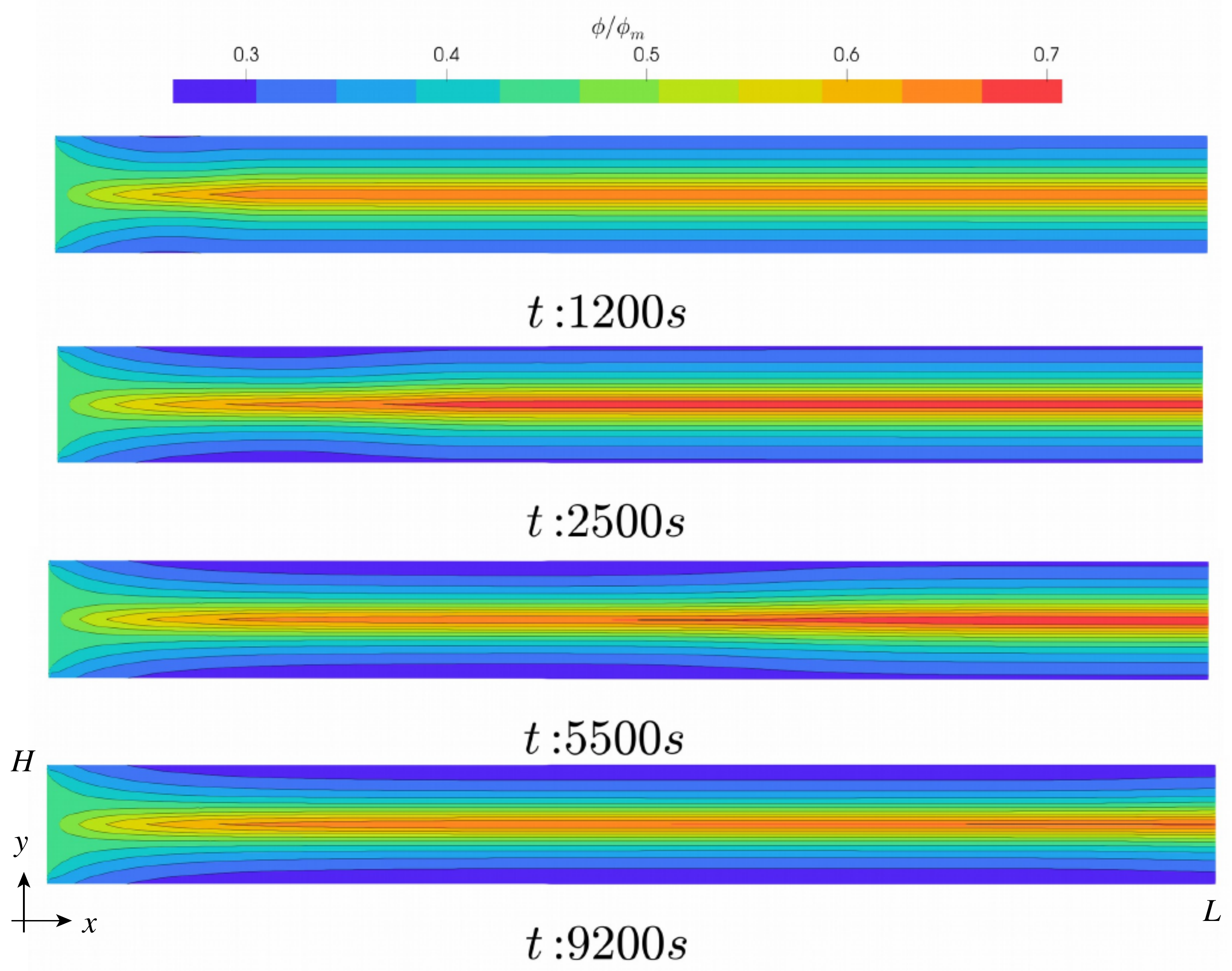}
 \caption{Time-evolution of the particle volume concentration in the ``full'' (non-periodic) 2D channel for $\phi_b = 0.3$. The channel has been scaled in the axial $x$-direction by a factor of $10^{-2}$ for display purposes.}
\label{fig::channelEvo}
\end{figure}

The disturbance induced by the inlet boundary condition is propagated at a finite speed throughout the domain, as shown in figure \ref{fig::channelEvo}. The time required for the inlet effects to propagate through the channel is significantly larger than the characteristic time of the particle migration process, and this should be considered when taking measurements in actual microchannels. In fact, far away from the inlet the concentration profile reaches an apparent steady state, which is analogous to our results from the fully periodic channel simulations in figure \ref{fig::channel}. Sections of the channel reached by the inlet disturbance switch from mass conservative (the area-averaged particle concentration is the same for each section) to flux conservative (the area-averaged particle concentration \emph{flux} is the same for each section). 

\subsection{Suspension flow in a cylindrical Couette cell}
\label{SS:R.2}

\begin{figure}
\centering
  \includegraphics[width=0.375\linewidth]{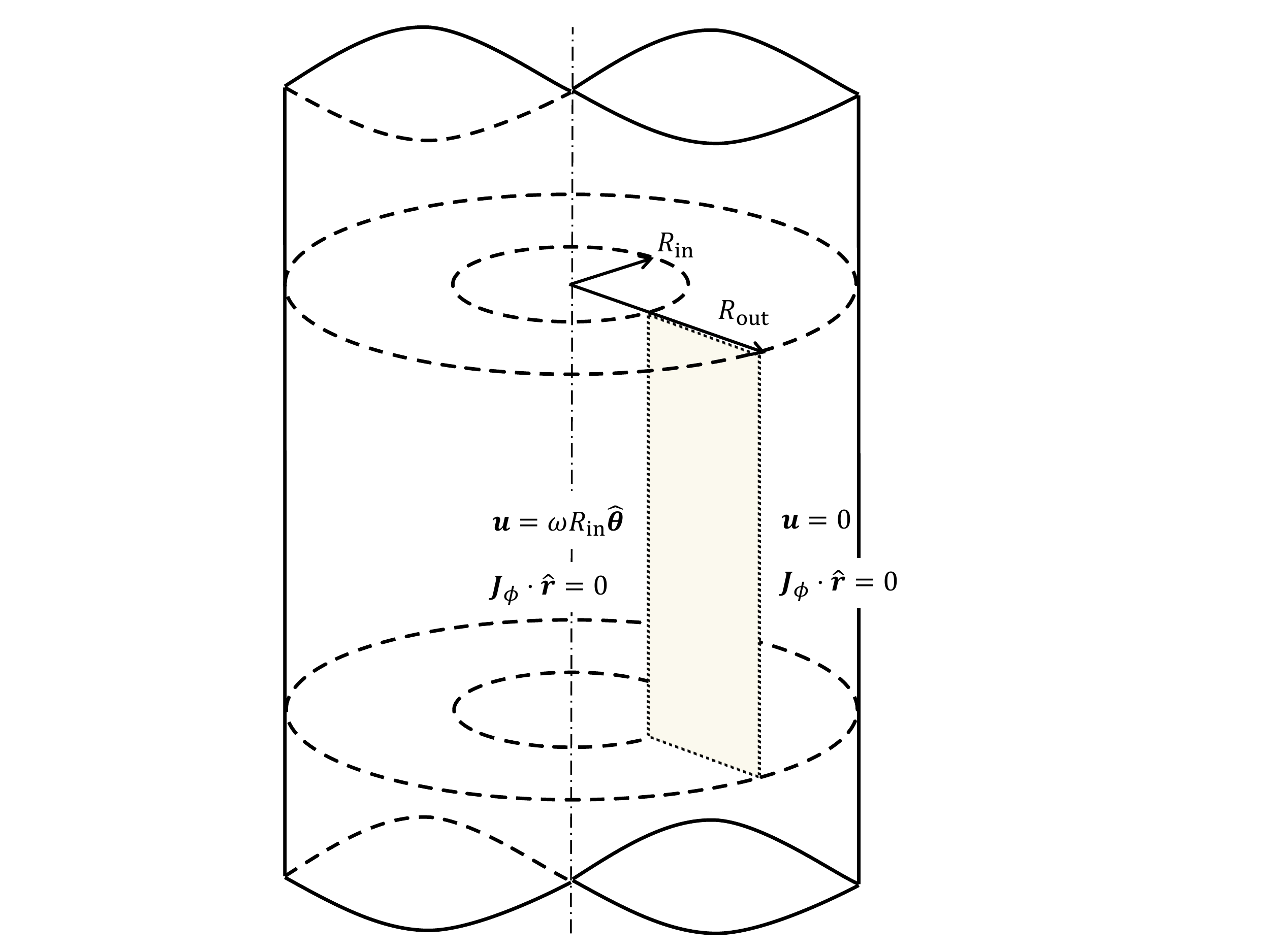}
 \caption{Schematic representation of the Couette cell geometry. The ``one-dimensional'' mesh corresponds to the shaded region in the radial direction between $r=R_\mathrm{in}$ and $r=R_\mathrm{out}$. Here, $\omega$ is the angular velocity of the inner cylinder, $\hat{\boldsymbol{\theta}}$ is the unit normal vector in the azimuthal direction, and $\hat{\bm{r}}$ is the unit normal vector in the radial direction. Boundary conditions are shown for the surfaces $r=R_\mathrm{in}$ and $r=R_\mathrm{out}$, while an empty boundary condition is applied in the axial direction. A `wedge' boundary condition is applied in the angular direction.}
\label{fig::CouetteScheme}
\end{figure}

Next, we test our TFM solver on the Couette cell geometry depicted in figure \ref{fig::CouetteScheme}, where the domain consists of the region between two concentric cylinders. The inner cylinder rotates about its central axis with angular velocity $\omega$, while the outer cylinder is fixed. The system is initialized with a uniform particle volume concentration $\phi_b=0.5$. The mesh consists of $20$ cells in the radial direction and the closure models are detailed in table \ref{tab::closuresCouette}. This flow configuration has been studied experimentally in \citep{pabga92} and numerically using the SBM in \citep{DLLM2013,MB1999}. Additionally, a semi-analytic model was proposed by Dbouk et al.~\citep{DLLM2013}. All these studies are in agreement and predict similar concentration profiles.

\begin{table}
    \centering
    \begin{tabular}{|c|c|l|}
    \hline
      closure  & expression & coefficients  \\
      \hline
        $f(\phi)$ & equation \eqref{eq::MorrisBoulay} & $\alpha=4$ \\
        $\eta_{\text{N}} (\phi)$ & equation \eqref{eq::etaN} & $K_{\text{N}}=0.75$ \\
        $\mu_p/\mu_f$ & equation \eqref{eq::closure_mup} & $a_{\mu} = 0$, $b_{\mu} = 2.5$, $c_{\mu} = 0.1$ \\
        $\lambda_i(\phi)$ & constant values & $\left\lbrace 1.0, \; 0.8, \; 0.5 \right\rbrace$ \\
        $\phi_m$ & constant value & $0.68$ \\
        $\dot{\gamma}_{\text{NL}}$ & constant value & $0$ \\
    \hline
    \end{tabular}
    \caption{Closure models and parameters used for the Couette (cylindrical) configuration.}
    \label{tab::closuresCouette}
\end{table}

\begin{figure}
\begin{subfigure}{.5\textwidth}
  \centering
  \includegraphics[width=0.95\linewidth]{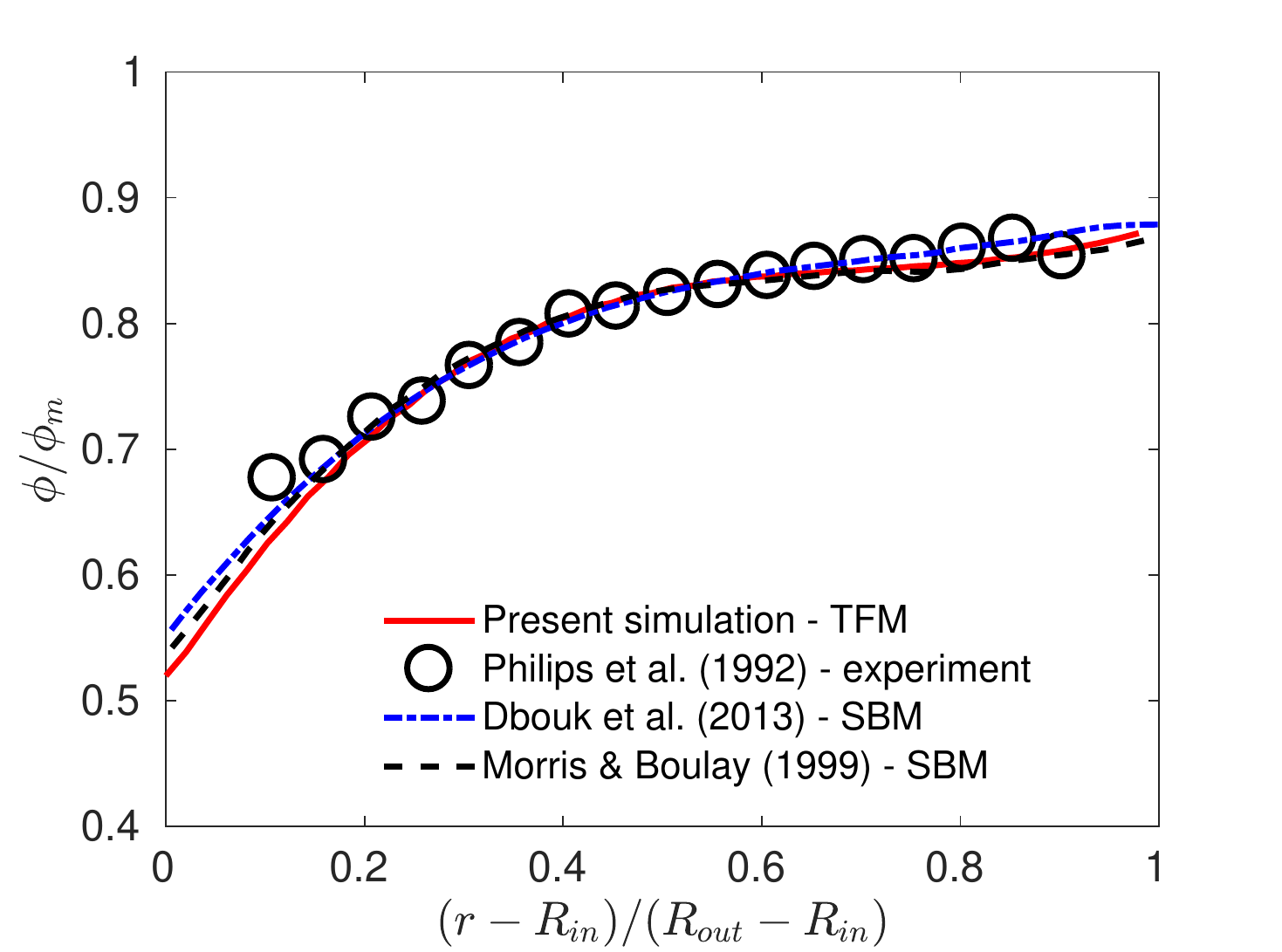}
  \caption{after $200$ turns}
  \label{fig:couette1}
\end{subfigure}%
\begin{subfigure}{.5\textwidth}
  \centering
  \includegraphics[width=0.95\linewidth]{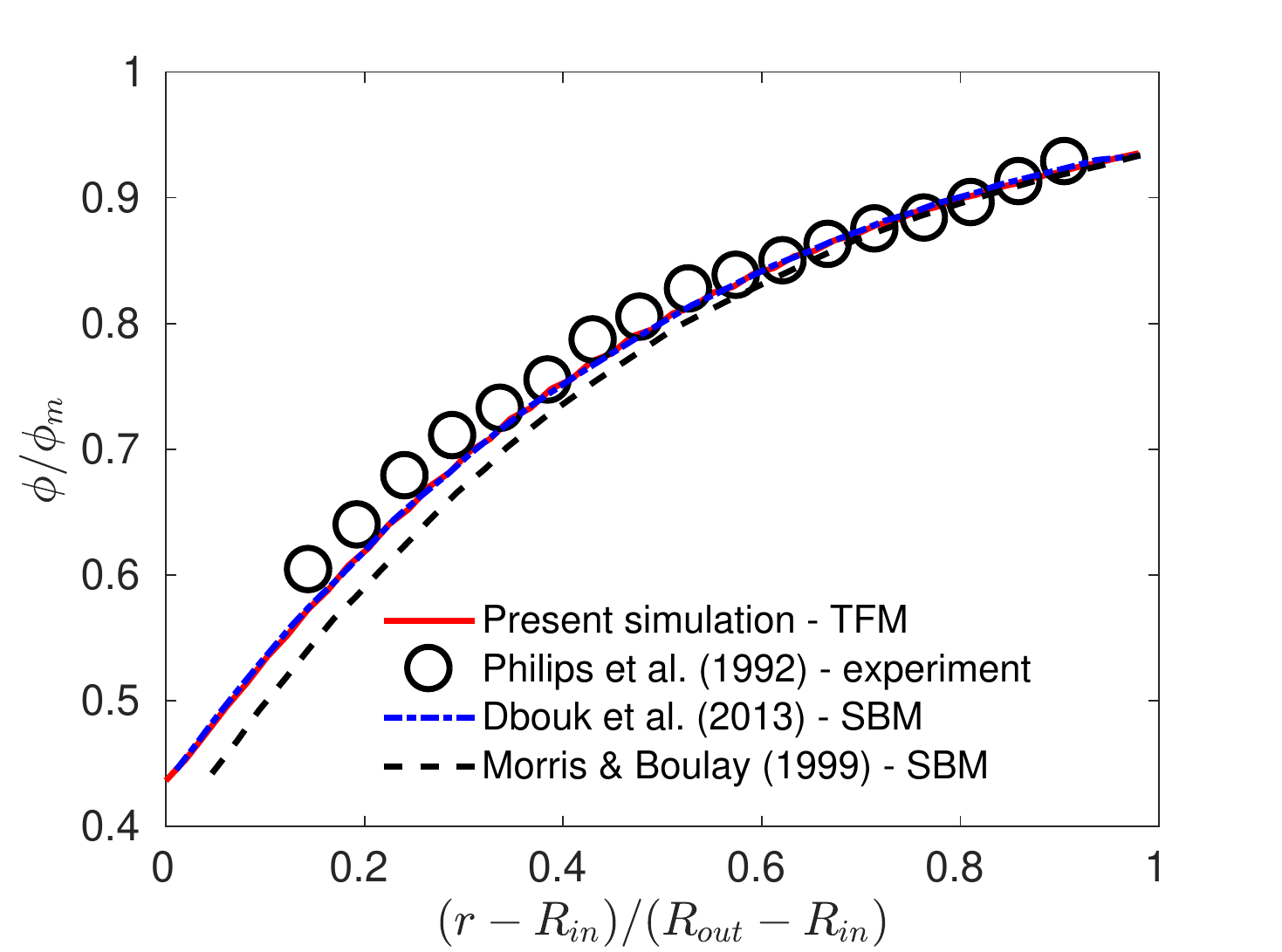}  \caption{after $12,000$ turns}
  \label{fig:couette2}
\end{subfigure}
\caption{Comparison between the two-fluid model (TFM), the suspension balance model (SBM) and experimental data for the Couette cell geometry. No significant deviation from the expected outcomes is observed for the TFM model, thus showing good agreement. }
\label{fig::couette}
\end{figure}

In the Couette cell, a homogeneous suspension fills the gap between two concentric cylinders of radii $R_\mathrm{in} = 0.64 ~\text{cm}$ and $R_\mathrm{out} = 2.34~\text{cm}$. At the initial time $t=0 ~\text{s}$, the inner cylinder starts spinning with angular velocity $\omega$, thus giving rise to a shear in the radial direction, which induces particle migration. Experiments employed a suspension composed of Poly(methyl methacrylate) (PMMA) spheres with mean diameter $d_p=675 ~\mu \text{m}$ suspended in a Newtonian fluid having dynamic viscosity $\mu_f = 9.45 ~\text{Pa s}$. The particle and fluid densities are $\rho_f = \rho_p = 1.183 ~\text{g} \, \text{cm}^{-3}$. We again employed the set of parameters suggested by Morris and Boulay \citep{MB1999} in our rheology models. Figure \ref{fig::couette} show that the TFM is in good agreement with results from previous studies.   

\subsubsection{Solver performance as $\phi\to\phi_m$}

Whenever the particle concentration approaches the maximum limit $\phi_m$, we are approaching the jamming limiting in which the flow arrests. The computational problem becomes very stiff due to the divergence of various quantities in the rheological closures above. It is certainly problematic to try to push a rheological model of flow into a jamming regime, the physics of which are dominated by particle re-arrangements and contact \cite{Morris2018}. Nevertheless, it might be illustrative to consider an example of pushing the proposed numerical method to its limit. In particular, it is important to know whether, in this situation, the algorithm may fail because the iteration loop diverges (no solution can be achieved).

Divergence and failure of the numerical algorithm can be avoided by approaching the jamming threshold $\phi_m$ smoothly and without sudden jumps. Therefore, one should:
\begin{itemize}
    \item[(i)] Employ the implicit formulation for the stress in the advection equation for the particle phase as detailed in section \ref{S:N}. This is particularly critical since the we observed that the algorithm diverges when using the standard explicit formulation approaching the jamming regime.
    \item[(ii)] Increase the number of pressure correctors to couple the momentum equations more tightly.
    \item[(iii)] Reduce the Courant number, which can also prevent sudden pressure spikes or instabilities.
\end{itemize}

\begin{figure}
    \centering
    \includegraphics[width=0.5\linewidth]{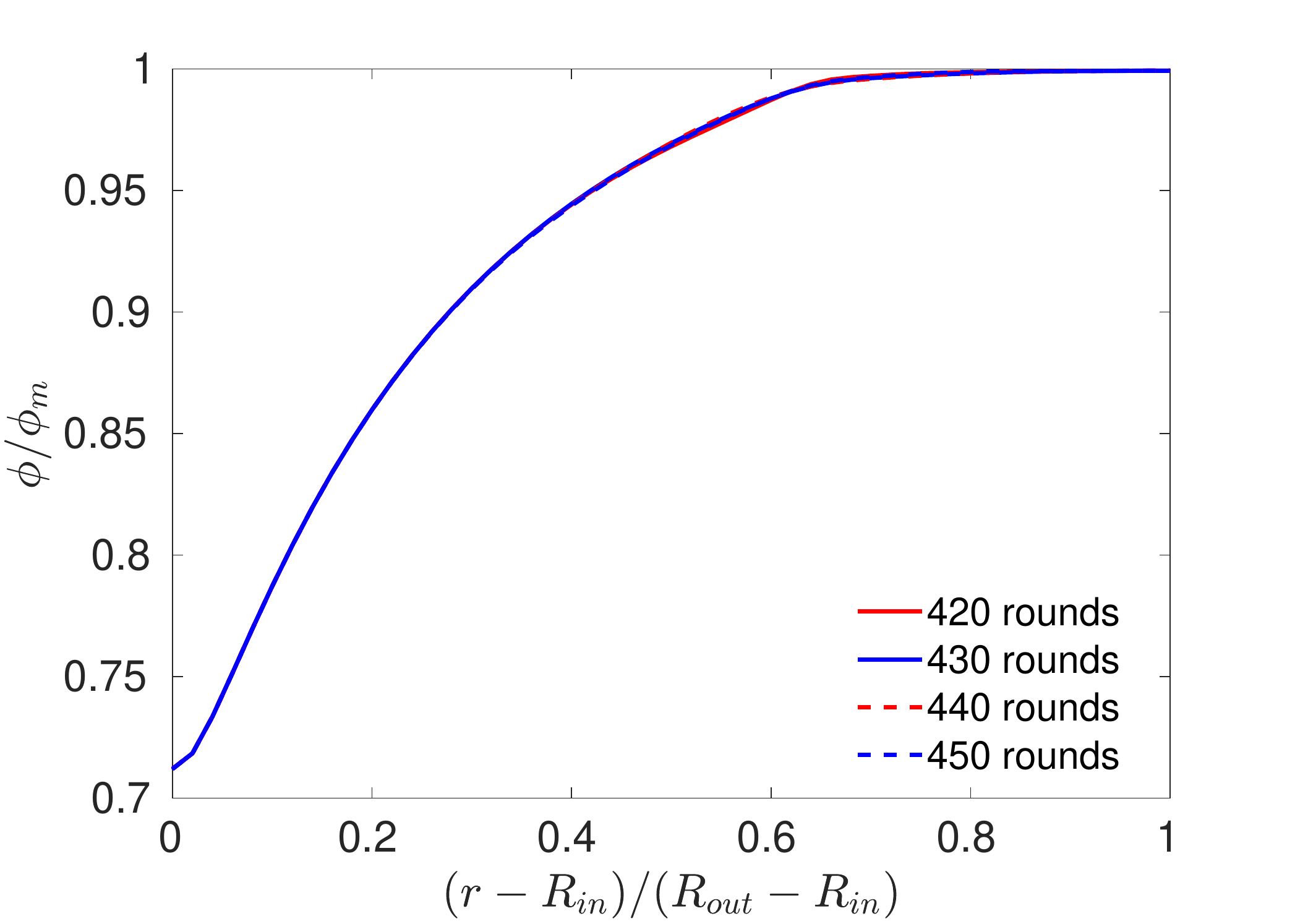}
    \caption{Particle concentration profiles in the jamming regime for a Couette cell. Different lines correspond to different simulation times, expressed in number of revolutions. Slight oscillations are noticeable in the region between dimensionless radii of $0.5$ and $0.7$.}
    \label{fig::jamming}
\end{figure}

Figure \ref{fig::jamming} shows that we observe some small oscillations int the Couette cell when approaching the jamming regime, if we employ (i)the implicit stress formulation, (ii) $20$ pressure correctors, and (iii) a maximum Courant number equal to $0.3$. The system was initialized with $\phi/\phi_m = 0.956$. Notice that the cells close to the origin have a smaller volume due to the cylindrical coordinate system employed and that mass was conserved during the whole simulation. 

\subsection{Resuspension}
\label{SS:R.3}

\begin{figure}
    \centering
    \includegraphics[width=0.6\linewidth]{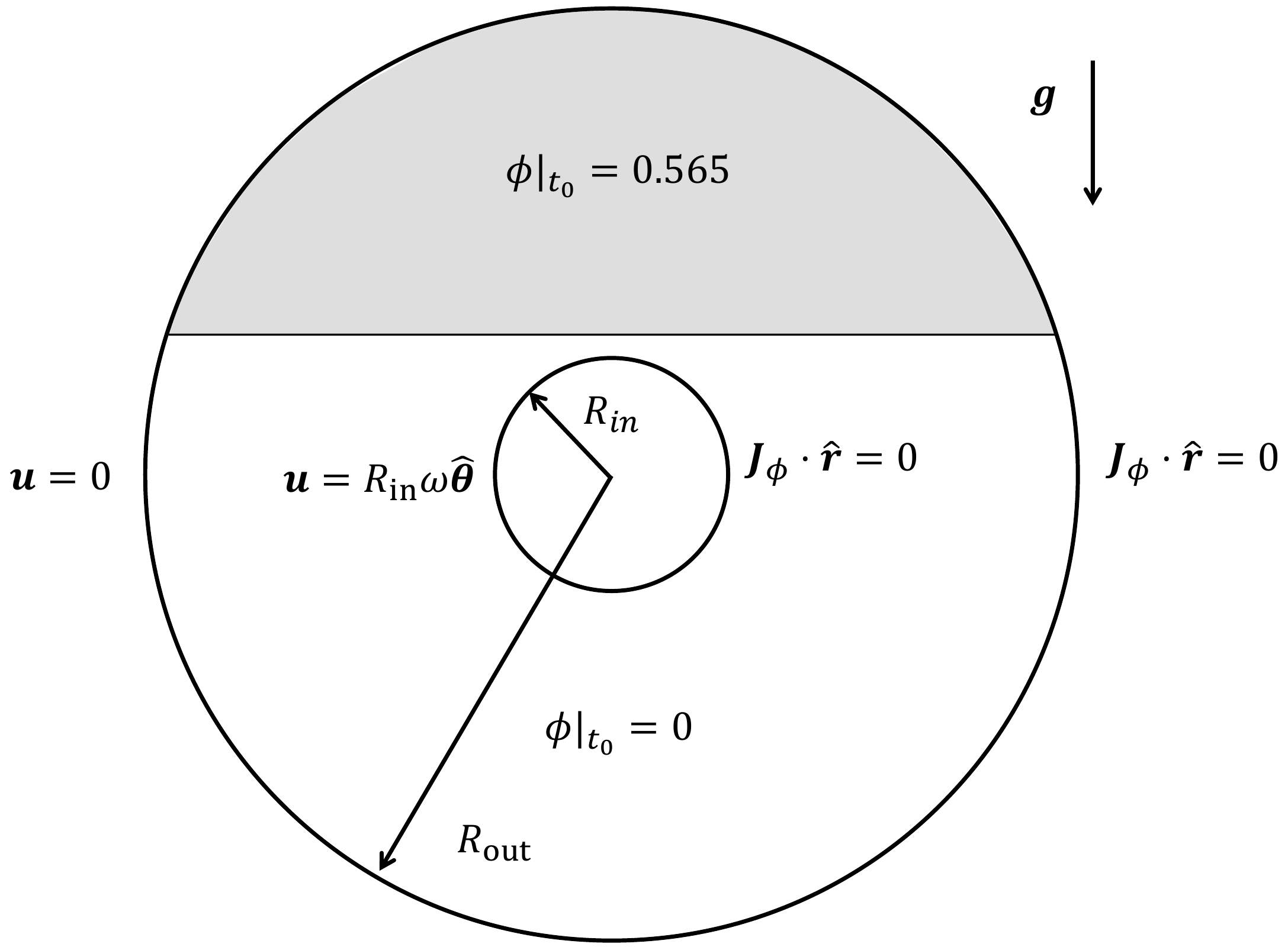}
    \caption{Schematic representation of a cylindrical mixer with rotating inner cylinder. The domain is initialized with two regions of uniform but different particle volume concentration $\phi|_{t_0}$, with clear fluid in the lower part of the mixer. The size of these regions is adjusted to obtain an average initial particle volume concentration $\phi_b=0.2$. Here, $\omega$ is the angular velocity of the inner cylinder, and $\hat{\boldsymbol{\theta}}$ is the unit normal vector in the azimuthal direction. Boundary conditions are shown for the surfaces $r=R_\mathrm{in}$ and $r=R_\mathrm{out}$, while an empty boundary condition is applied in the axial direction.}
    \label{fig::resuspensionScheme}
\end{figure}

Our next validation test for the TFM is in a curvilinear mixing flow with buoyancy effects, as shown schematically in figure \ref{fig::resuspensionScheme}. This benchmark is also an important test of the frame-invariant form of the anisotropic tensor $\bm{Q}$ that we introduced in equations~\eqref{eq::Q} and \eqref{eq::unitVectors_Q} above. This flow configuration was first employed by Abbot et al.~\citep{ATG1991} to investigate particle migration, and it has been a staple in subsequent experimental and numerical works \citep{RLS2002,DLLM2013}. 

A suspension of particles with diameter $d_p = 494 ~\mu \text{m}$ and density $\rho_p = 1.18 ~\text{g} \, \text{cm}^{-3}$ is suspended in a Newtonian fluid with density  $\rho_f = 1.253 ~\text{g} \, \text{cm}^{-3}$ and viscosity $\mu_f = 0.588 ~\text{Pa s}$. The suspension fills the gap between two concentric cylinders of inner radius $R_\mathrm{in}=0.64 ~ \text{cm}$ and outer radius $R_\mathrm{out}=2.54 ~\text{cm}$. The inner cylinder is set into motion by rotating it anti-clockwise, which shears the fluid, introducing a velocity gradient in the radial direction. 

The fully structured mesh consists on four blocks of $50\times50$ cells, resulting in $50$ cells in the radial direction and $200$ cells in the angular direction. For time stepping, a maximum Courant number of $0.4$ was imposed. This value was chosen to allow the solver to finish the \emph{PIMPLE} loop in a flow configuration having a Courant number higher than the maximum, while still fulfilling the Courant--Friedrich--Lewy (CFL) condition \cite{CFL1928}. In fact, the CFL condition tends to be violated near the inner cylinder when imposing a maximum Courant number close to one, as can be seen in the results of Dbouk et al.~\citep{DLLM2013}. The closure models employed for this benchmark are detailed in table \ref{tab::closuresResuspension}.

\begin{table}[b]
    \centering
    \begin{tabular}{|c|c|l|}
    \hline
      closure  & expression & coefficients  \\
      \hline
        $f(\phi)$ & equation \eqref{eq::MillerMorris} & $\alpha=4$ \\
        $\eta_{\text{N}} (\phi)$ & equation \eqref{eq::etaN} & $K_{\text{N}}=0.75$ \\
        $\mu_p/\mu_f$ & equation \eqref{eq::closure_mup} & $a_{\mu} = -1$, $b_{\mu} = 0$, $c_{\mu} = 1$ \\
        $\lambda_i(\phi)$ & equation \eqref{eq::lambda_DLL} & -- \\
        $\phi_m$ & constant value & $0.64$ \\
        $\dot{\gamma}_{\text{NL}}$ & constant value & $0$ \\
    \hline
    \end{tabular}
    \caption{Closure models and parameters used in the resuspension configuration.}
    \label{tab::closuresResuspension}
\end{table}

\begin{figure}[ht!]
    \centering
    \includegraphics[width=\linewidth]{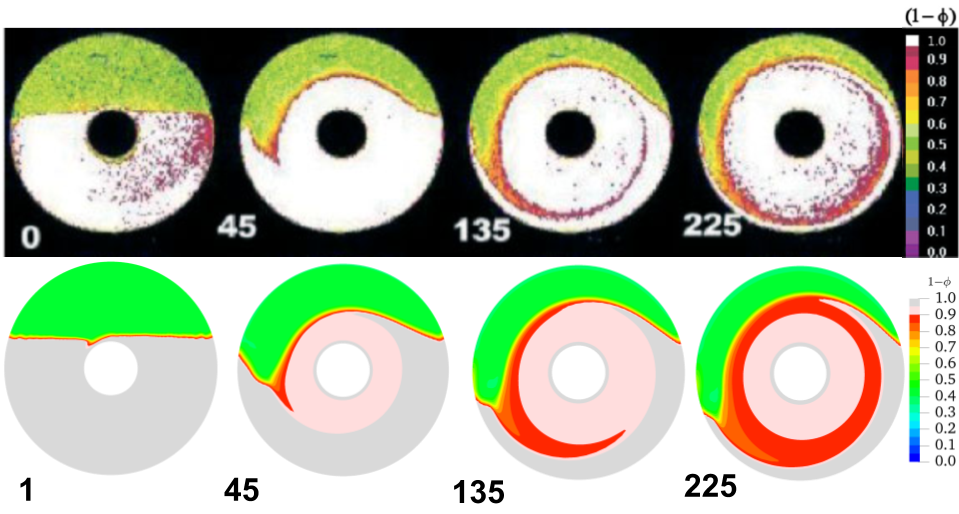}
    \caption{Comparison between our numerical TFM results (bottom) and the resuspension experiment from Rao et al.~\cite{RLS2002} (top) [Reprinted with permission from \cite{RLS2002} \copyright 2002 John Wiley \& Sons, Ltd]. Numbers on the bottom left correspond to the number of turns of the inner cylinder.}
    \label{fig::resuspension}
\end{figure}

As shown in figure \ref{fig::resuspension}, our TFM implementation is able to capture a range of features in this flow, such as the formation of a thin particle layer in the mixing direction as well as the existence of a low particle density region at the bottom of the cylinder. However, we point out that the particle volume fraction distribution is strongly dependent on the choice of the closure models and the corresponding closure coefficients \citep{DLLM2013}, in particular the choice of $\phi_m$. This sensitivity means that model calibration is needed when simulating such complex flows. Therefore, future research should address the issue of `universal' rheological closures for dense suspensions.

\subsection{Secondary flows: Symmetric herringbone channel}

\begin{figure}[h]
    \begin{subfigure}{0.49\linewidth}
        \centering
        \includegraphics[height=0.55\linewidth]{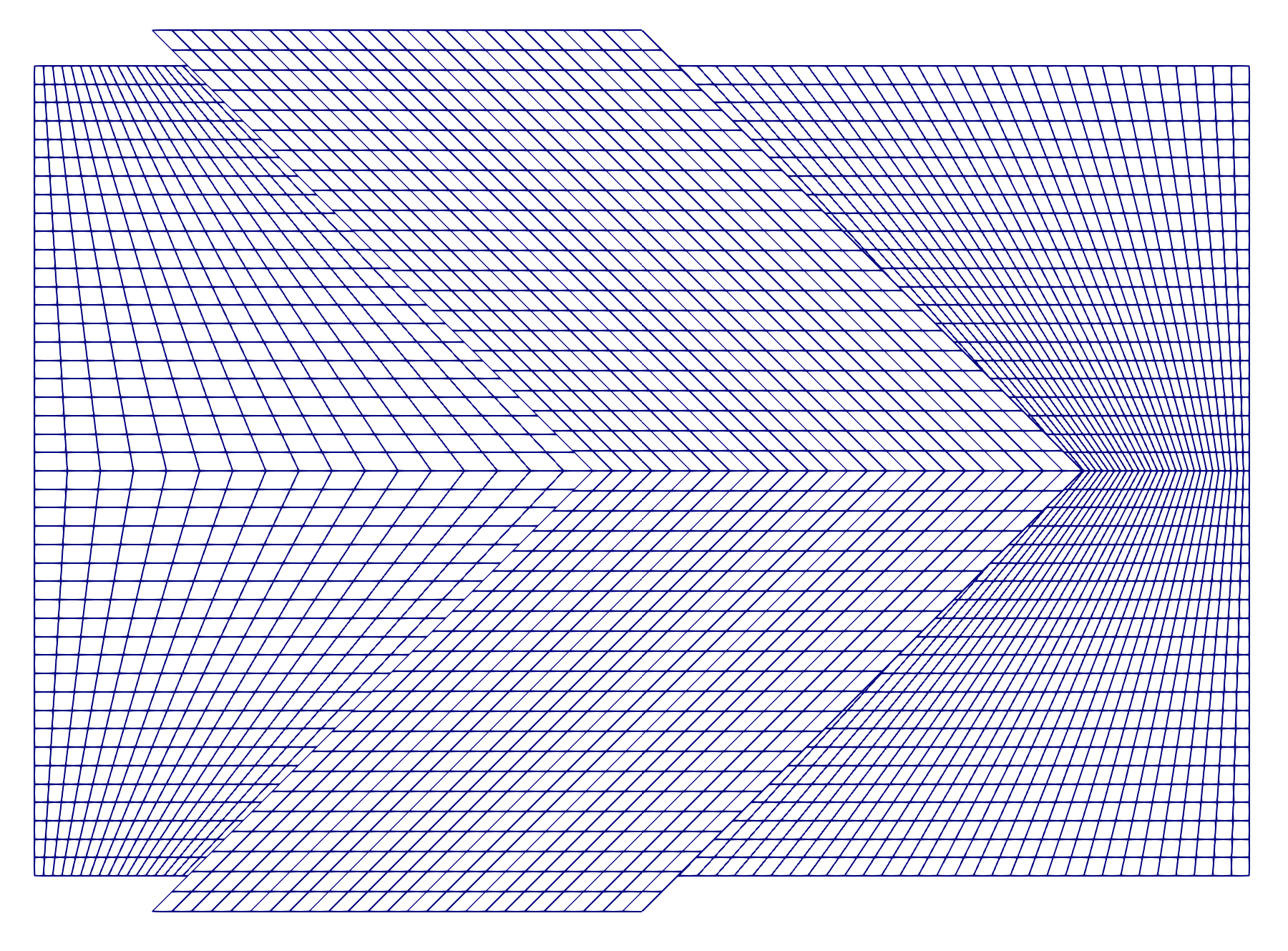}
        \caption{Top view}
        \label{fig::HBtop}
    \end{subfigure}
    \begin{subfigure}{0.49\linewidth}
        \centering
        \includegraphics[height=0.55\linewidth]{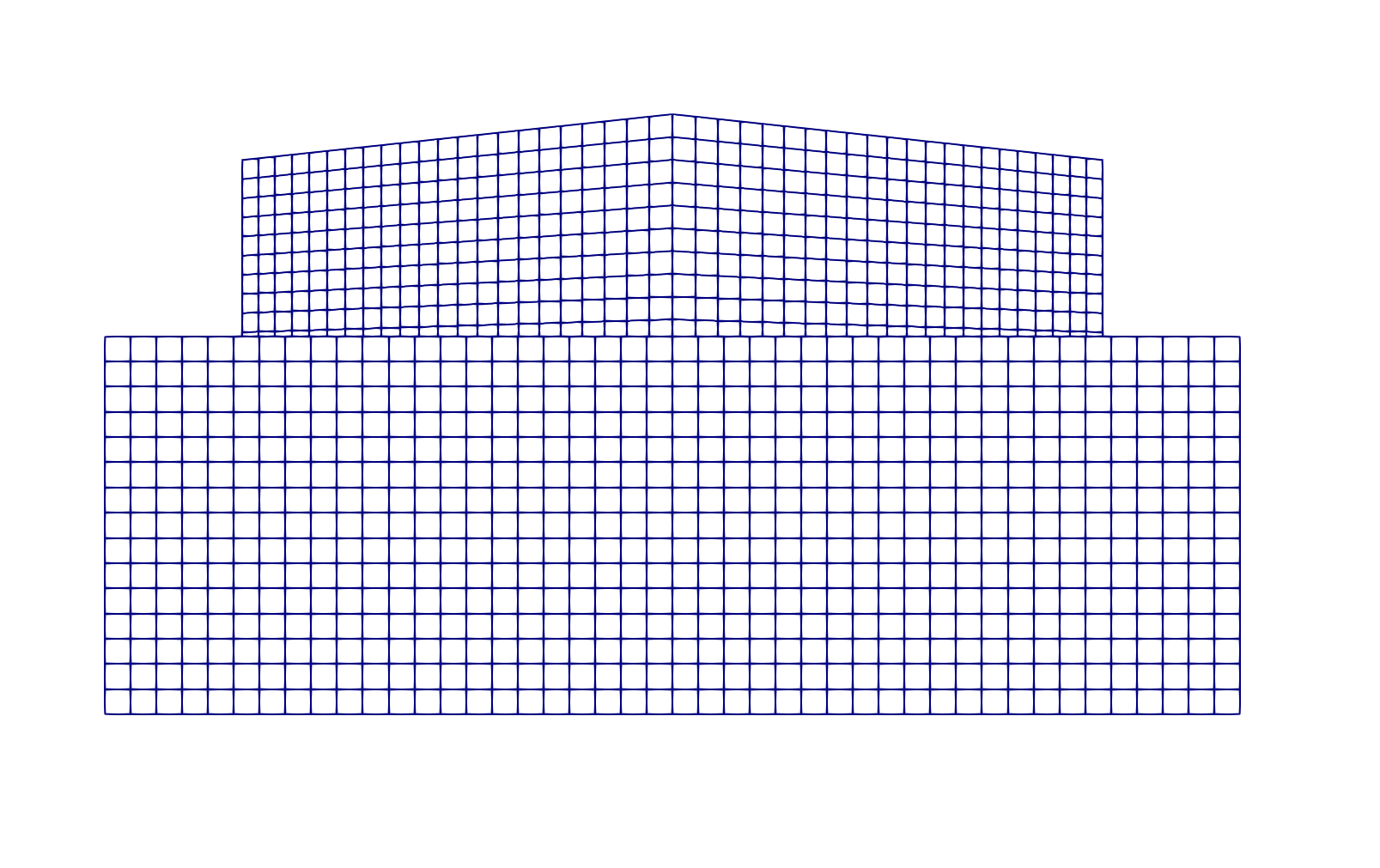}
        \caption{Front view}
        \label{fig::HBfront}
    \end{subfigure}
    \caption{Computational grid employed for the simulation of secondary flows in a symmetric herringbone channel. A square channel with dimensions $135 ~\mu\text{m}\times 90\mu\text{m} \times 30 ~\mu\text{m}$ is employed and the herringbone structure is $20 ~\mu\text{m}$ high and $50 ~\mu \text{m}$ long.}
    \label{fig::HB_grid}
\end{figure}

Suspension flows in channels with 2D and 3D flows features have been studied experimentally \cite{GG2008,GXG09} due to their importance for enhancing mixing and transport rates at low Reynolds numbers \cite{OW04}. Specifically, symmetric herringbone channels  (inspired by the so-called ``staggered herringbone mixer'' \cite{Stroock02}) lead to the emergence of a vertical band of low concentration in the center of the channel, thus a particle migration flux is established towards the lateral walls. In this subsection, we simulate this phenomenon using the proposed TFM. The geometry employed is depicted in figure \ref{fig::HB_grid} together with the numerical grid. The geometry and material properties are chosen according to the experiment of Gao and Gilchrist \cite{GG2008}. 

We employed a fully periodic domain, and we enforced the flow rate via a forcing term in the governing equations. Additionally, we initialized the system with a uniform suspension with particle volume concentration $\phi_b = 0.1$. The choice of a fully periodic domain induces a significant difference with respect to experimental works, were a suspension was pumped in a long channel initialized with a clear fluid. In fact, not only do we expect a similar discrepancy as that discussed in section~\ref{SS:R.1}, but we also do not expect that the Kelvin--Helmoltz (KH) instability discussed in \cite{GG2008} would arise, as this instability is due  to the shearing of the clear fluid initially filling the cavities (while we have initialized our simulations with a homogeneous suspension therein). The KH instability latter effect is responsible for the asymmetric concentration profile observed in the experiments \cite{GG2008}.

For these simulations, we employed particles with diameter $d_p = 1.01 ~\mu \text{m}$ and density $\rho_p = 2 ~\text{g/cm}^3$, while the fluid has density $\rho_f = 1.2~ \text{g}\,\text{cm}^{-3}$ and viscosity $\mu_f = 0.04 ~\text{Pa s}$. Thus, particles will tend to sink due to the density difference. The closures employed in this simulation are the same as in table \ref{tab::closuresPoiseuille}. Due to the non-orthogonality of the mesh, multiple corrector steps are employed to obtain a stable solution.

\begin{figure}[ht]
    \begin{subfigure}{0.49\linewidth}
        \centering
        \includegraphics[width=\linewidth]{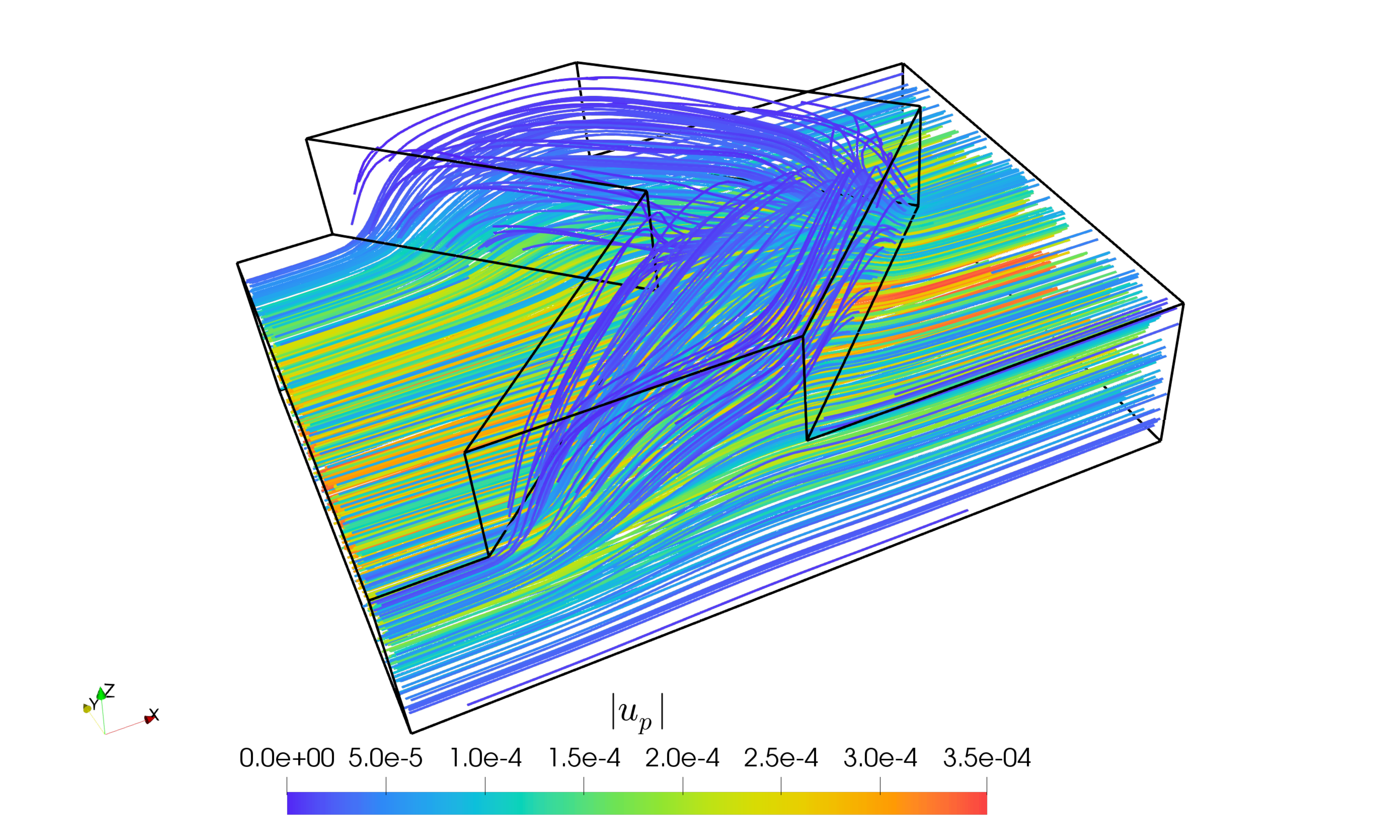}
        \caption{}
        \label{fig::HBvel}
    \end{subfigure}
    \begin{subfigure}{0.49\linewidth}
        \centering
        \includegraphics[width=\linewidth]{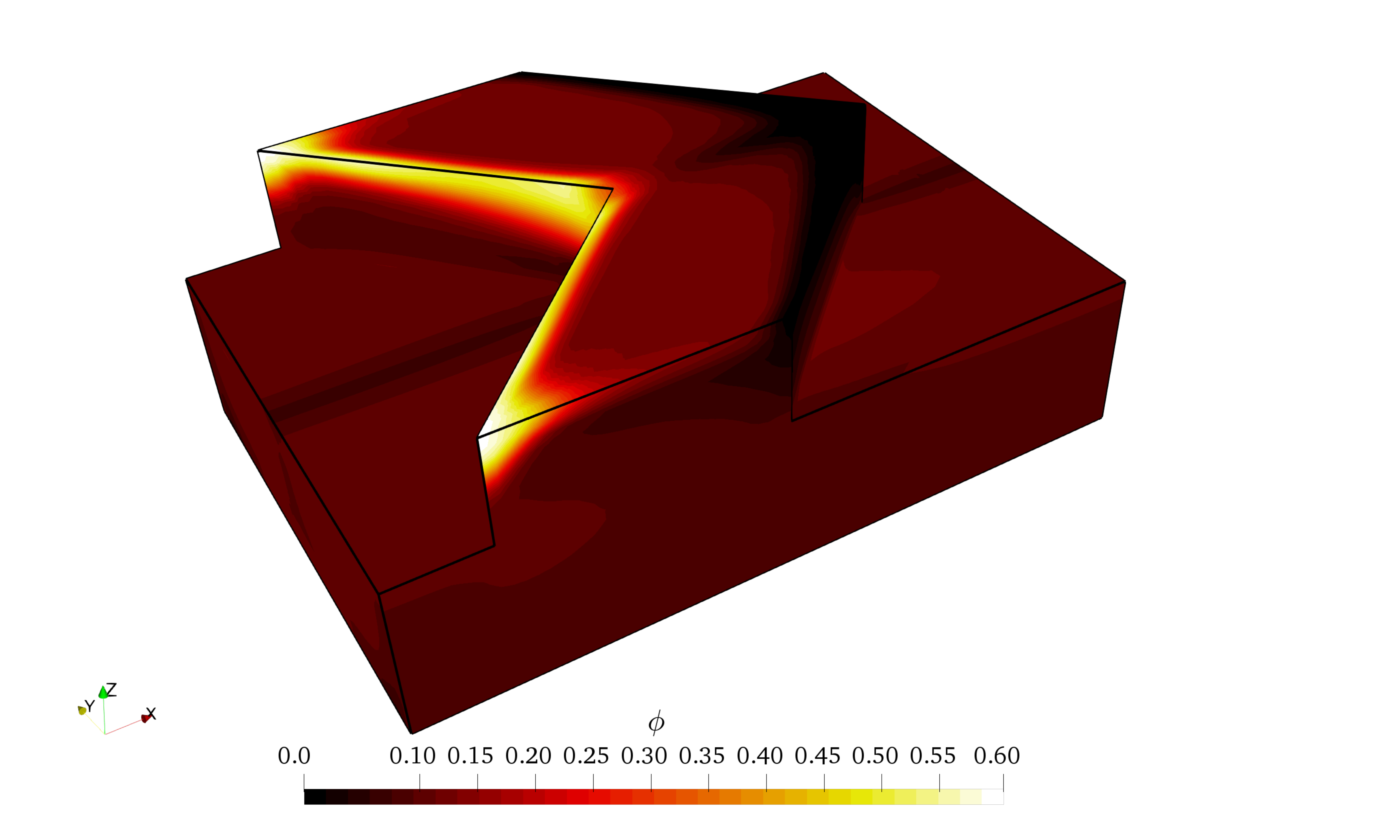}
        \caption{}
        \label{fig::HBphi}
    \end{subfigure}
    \caption{(a) Particle velocity streamlines and (b) particle volume concentration in the symmetric herringbone channel computed using the TFM. }
    \label{fig::HBfields}
\end{figure}

Figure \ref{fig::HBfields} shows the velocity and particle volume concentration after a steady state is reached. We observe that the particle volume concentration reaches its maximum and minimum values inside the cavity, corresponding to the most and least quiescent regions. In fact, we observed that, during the first few time steps,  the particle concentration increases in the vertical direction in the channel and subsequently decreases while particles are accumulating in the rear of the herringbone structure. This phenomenon would be less evident in experiments due to the presence of clear fluid in the cavities.

\begin{figure}
    \centering
    \begin{subfigure}{0.5\linewidth}
        \centering
        \includegraphics[width=0.9\linewidth]{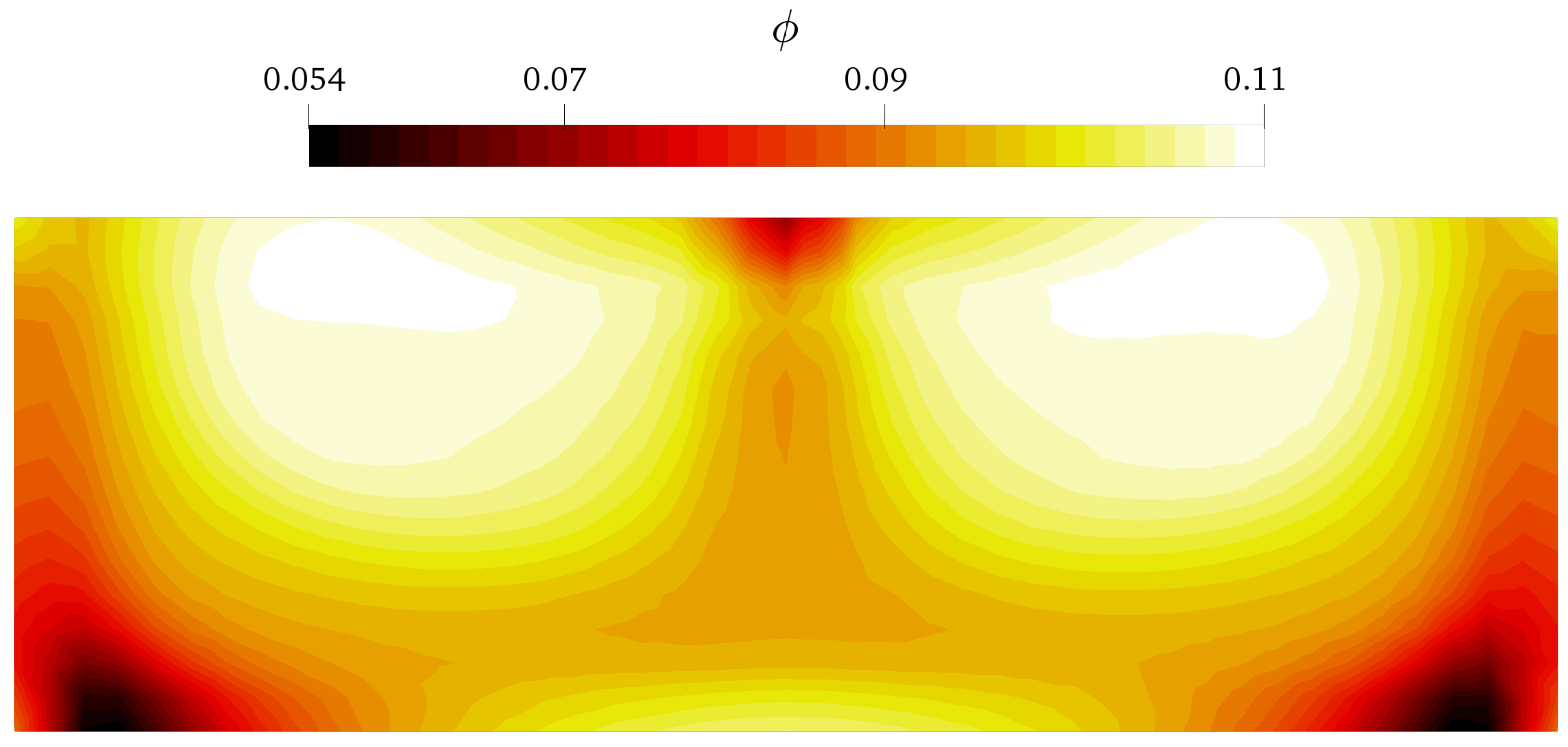}
        \vspace{5mm}
        \caption{}
        \label{fig::HBsec}
    \end{subfigure}%
    \begin{subfigure}{0.5\linewidth}
        \centering
        \includegraphics[width=\linewidth]{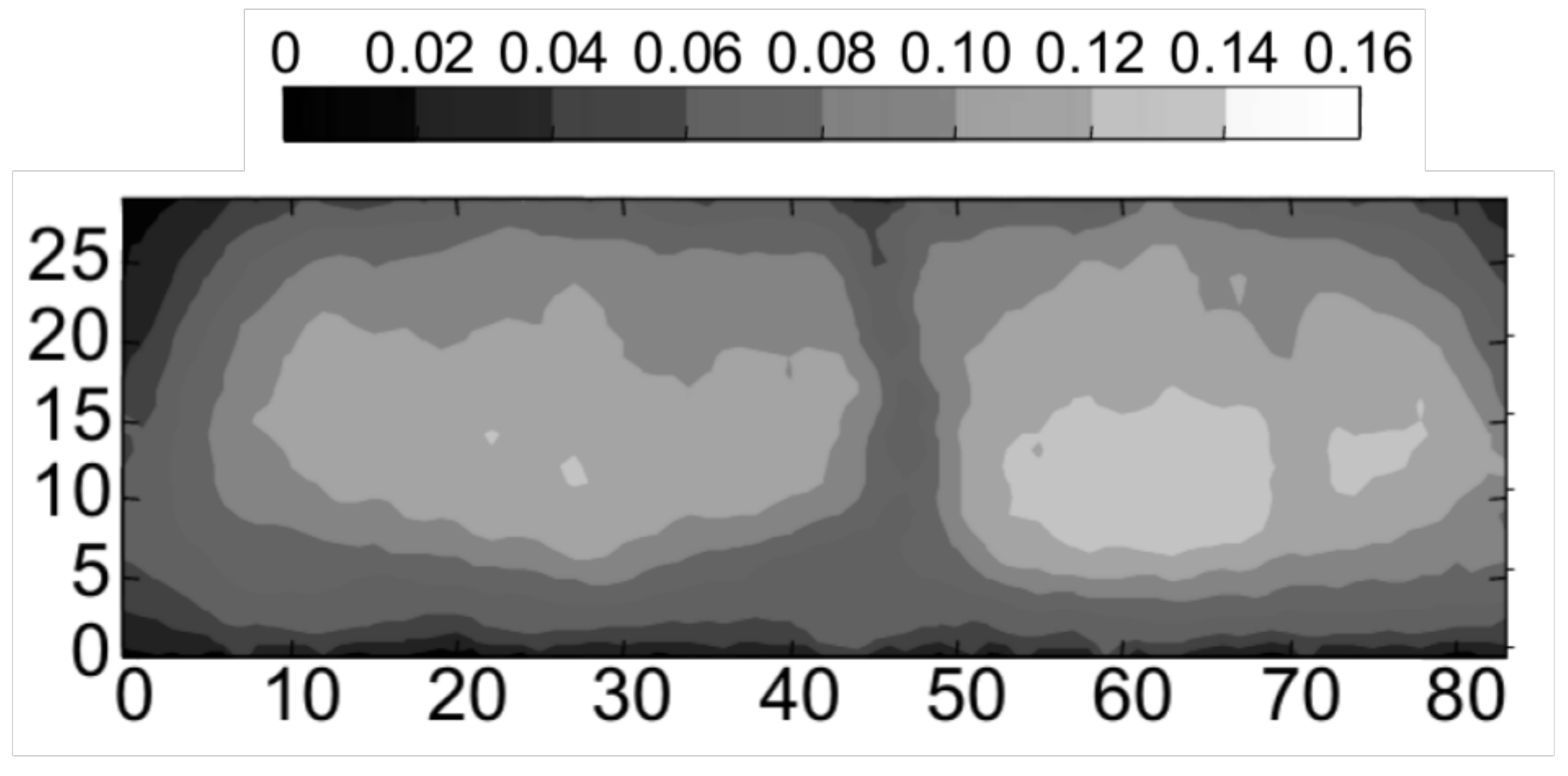}
        \caption{}
        \label{fig::HBsec2}
    \end{subfigure}
    \caption{Comparison between (a) the TFM solver and (b) the experiment from Gao and Gilchrist~\cite{GG2008} [Reprinted with permission from \cite{GG2008} \copyright 2008 
American Physical Society]. Color maps represent the particle volume concentration.}
    \label{fig::HBCompare}
\end{figure}

A comparison between our results and those from Gao and Gilchrist \cite{GG2008} is shown in figure \ref{fig::HBCompare}. Clearly, our TFM solver is able to predict the existence of two symmetric regions separated by a vertical line of low particle concentration. Obtaining quantitative agreement through extensive model calibration is beyond the scope of this simulation, since that would also require much more faithful numerical modeling of the experimental conditions. However, this test case demonstrates that our TFM solver is able to reproduce the physics of particle migration  induced by complex features in the flow geometry.

\section{Conclusions}
\label{S:C}
In this work, we presented a two-fluid formulation of the shear-dominated flow of a dense suspension. The proposed two-fluid model (TFM) allows us to simulate general unsteady curvilinear flows accounting for anisotropic constitutive models.  The TFM was implemented as an extension of the OpenFOAM{\textsuperscript\textregistered} \emph{twoPhaseEulerFoam} solver, and it is thus freely available for usage and improvement by anyone. We demonstrated that the solver is capable of accurately reproducing results from experiments and previous simulations based on the (less general) suspension balance model. Furthermore, the solver can be employed to study complex curvilinear suspension flows, and it can handle various non-orthogonal geometries. Therefore, in future work, the proposed TFM could be adapted to study, for example, highly unsteady particle migration in oscillatory flows in cylindrical geometries, a topic of significant current interest \cite{SBG16,CHMT17}, and provide further insight into Taylor dispersion of dense suspensions \cite{gs12,R13}.

However, further research needs to be performed in order to develop rheological models tailored for the TFM. Specifically, models that distinguish between long-range hydrodynamic interactions \cite{Morris2018} and contact frictional forces should be researched in order to develop a simulation tool that is capable of predicting transitions between flow regimes (for example, from non-Brownian to Brownian). Therefore, future research should perhaps be devoted to isolating the rheology of the particle phase from that of the mixture.  

The OpenFOAM{\textsuperscript\textregistered} code associated with this work can be freely downloaded from the first author's GitHub repository (\url{https://github.com/fmuni/twoFluidsNBSuspensionFoam}), together with the cases corresponding to the simulations performed in this work.

\section*{Acknowledgements}
Acknowledgment is made to the donors of the American Chemical Society Petroleum Research Fund for support of this research under ACS PRF award \# 57371-DNI9. P.P.N.\ additionally acknowledges the S.N.\ Bose Scholars program of SERB-IUSSTF for funding his internship at Purdue during Summer 2018. We thank the two anonymous reviewers for comments and questions that have improved the manuscript.

\appendix
\label{appendix:A}
\section{Grid convergence analysis}

In this appendix, we show that the numerical grids we employed are sufficient to capture the requisite details of the suspension flows considered.  We express the degree of refinement using the number of cells $h$ in the shear direction, which is defined as $h = H/\Delta y$ for the parallel channel and as $h = (R_\mathrm{out} - R_\mathrm{in})/\Delta r$ for the Couette cell.

\begin{figure}
\begin{subfigure}{.5\textwidth}
  \centering
  \includegraphics[width=\linewidth]{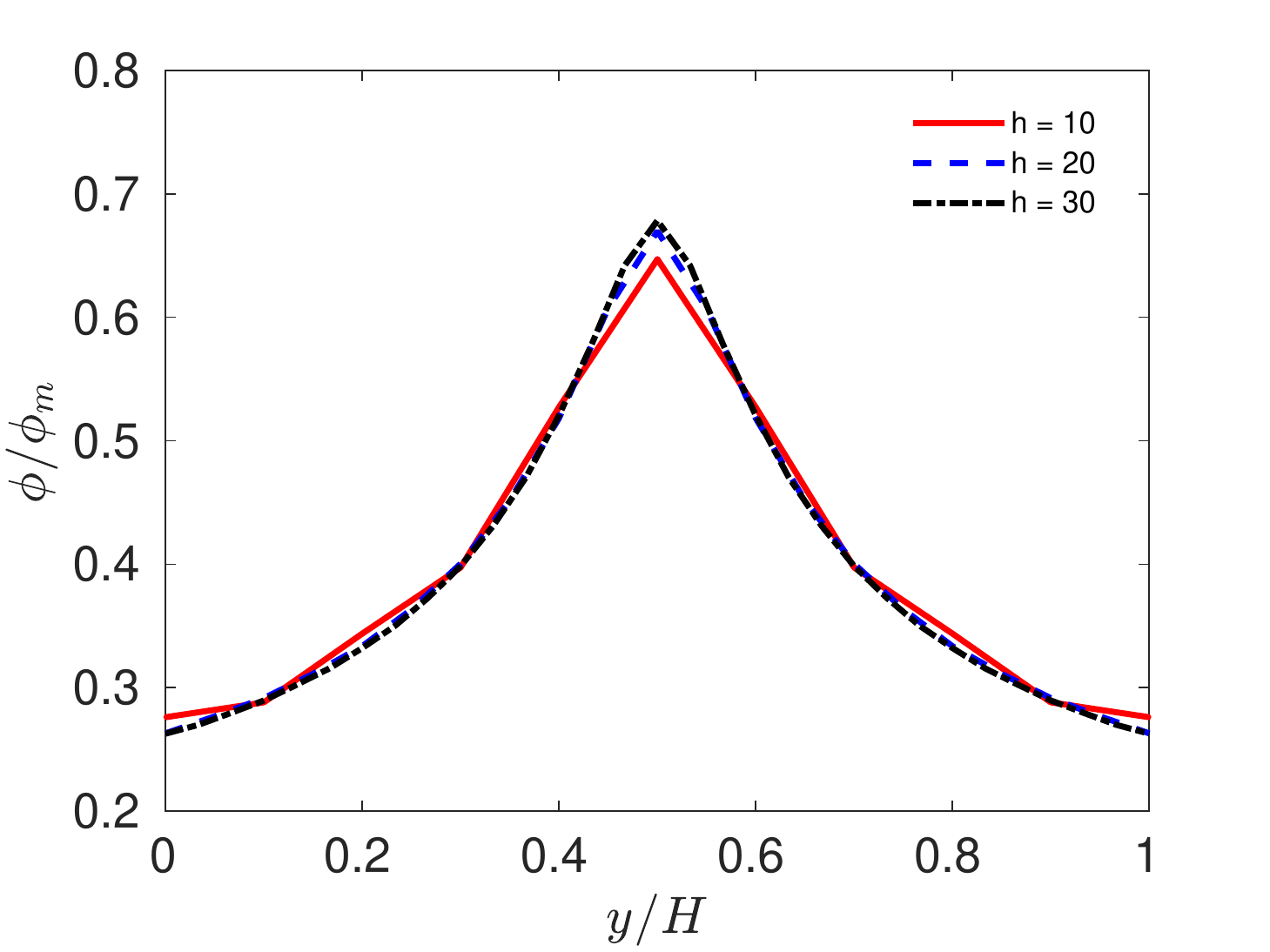}
  \caption{}
  \label{fig::gridConvergence_a}
\end{subfigure}%
\begin{subfigure}{.5\textwidth}
  \centering
  \includegraphics[width=\linewidth]{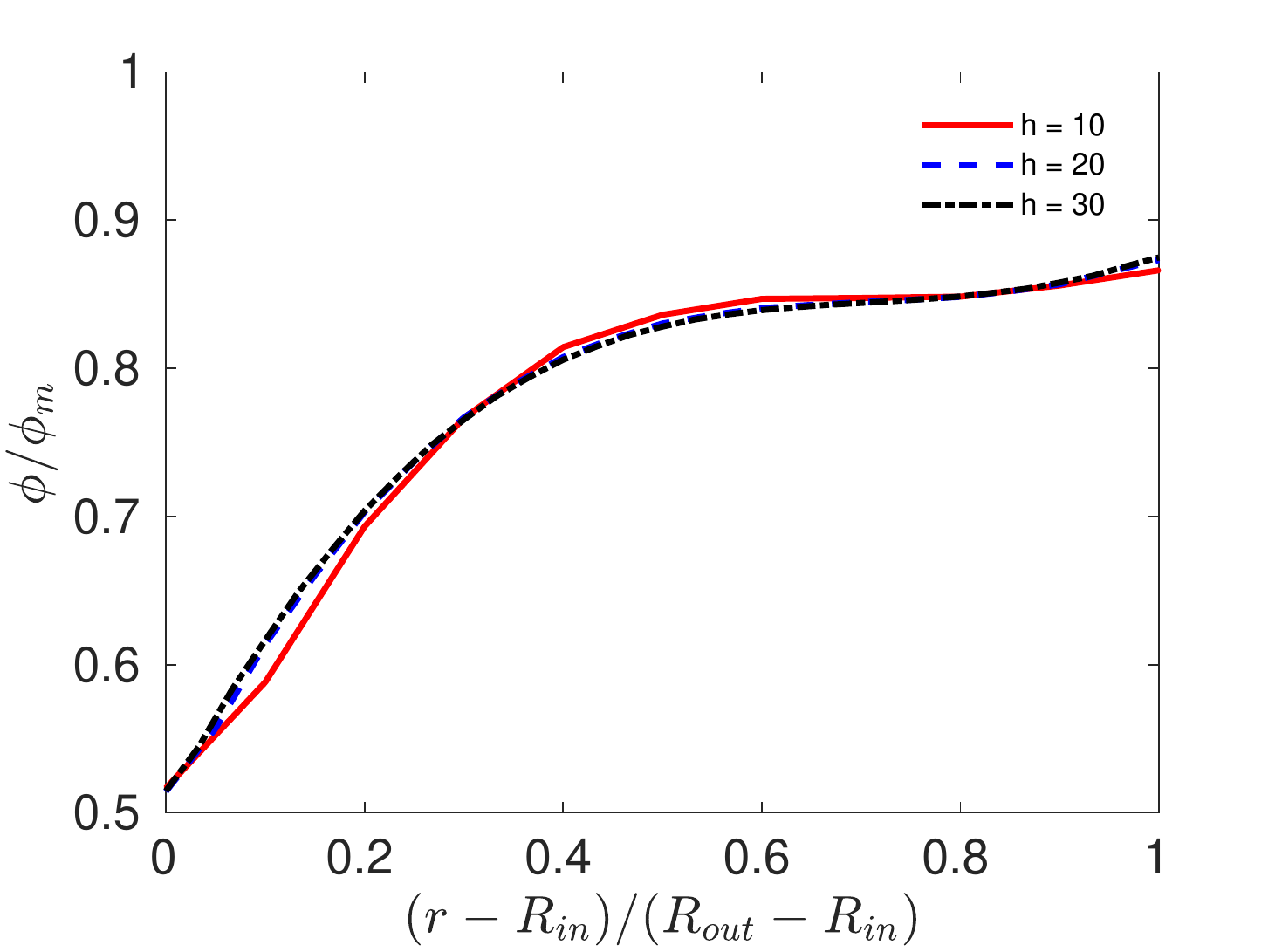} 
  \caption{}
  \label{fig::gridConvergence_b}
\end{subfigure}
\caption{Particle volume concentration profiles obtained using different grid resolutions for (a) the Pouiseille channel flow ($\phi_b=0.3$) and (b) the Couette cell flow ($200$ turns of the inner cylinder).}
\label{fig::gridConvergence}
\end{figure}

Concerning the Poiseuille channel flow, figure \ref{fig::gridConvergence_a} shows the case of $\phi_b = 0.3$, which is the most sensitive to the grid size (since the particle volume concentration is far from the packing fraction $\phi_m$). We remark that using different expressions for the non-local shear rate $\dot{\gamma}_\mathrm{NL}$ leads to different grid dependencies, since the non-local shear rate is meant to limit the particle volume concentration at the centerline of the channel.

Similarly, figure~\ref{fig::gridConvergence_b} shows that, in the case of the Couette cell flow, no significant dependence on the grid size is observed for grids finer than $h = 20$. A non-local shear rate was not employed in this case.

\bibliographystyle{elsarticle-num-alpha}

\bibliography{references}


\end{document}